\pdfoutput=1
\documentclass[twocolumn,english,aps,superscriptaddress,prb]{revtex4-1}

\usepackage{babel}
\usepackage{amsmath}
\usepackage{amssymb}
\usepackage{graphicx}
\usepackage{xcolor}
\usepackage{physics}
\usepackage{comment}
\usepackage{subfig}
\usepackage{floatrow}
\usepackage{simpler-wick}
\usepackage[colorlinks,bookmarks=true,citecolor=blue,linkcolor=blue,urlcolor=blue, breaklinks=true]{hyperref}
\usepackage[labelsep=period]{caption}

\begin{document}

\title{Emergent U(1) symmetry in non-particle-conserving one-dimensional models}
\renewcommand{\figurename}{FIG.}

\author{Zakaria Jouini}
\affiliation{Institute of Physics, Ecole Polytechnique F\'ed\'erale de Lausanne (EPFL), CH-1015 Lausanne, Switzerland}
\author{Natalia Chepiga}
\affiliation{Kavli Institute of Nanoscience, Delft University of Technology, Lorentzweg 1, 2628 CJ Delft, The Netherlands}
\author{Loïc Herviou}
\affiliation{Institute of Physics, Ecole Polytechnique F\'ed\'erale de Lausanne (EPFL), CH-1015 Lausanne, Switzerland}
\author{Fr\'ed\'eric Mila}
\affiliation{Institute of Physics, Ecole Polytechnique F\'ed\'erale de Lausanne (EPFL), CH-1015 Lausanne, Switzerland}

\date{\today}
\begin{abstract} 
The properties of stable Luttinger liquid phases in models with a non-conserved number of particles are investigated. We study the Luttinger liquid phases in one-dimensional models of hard-core boson and spinless fermion chains where particles can be created and annihilated three by three on adjacent sites. We provide an intuitive and systematic method based on the flow equation approach, which accounts for additional terms in the correlations generated by the $\mathbb{Z}_3$-symmetric interactions. We find that despite the emergence of U(1) symmetry under renormalization, the observables are still affected by its breaking in the bare Hamiltonian. In particular, the standard bosonization mapping becomes insufficient to capture the full behavior of correlation functions.
\end{abstract}

\maketitle

\section{Introduction} 
The observation of density-wave order in recent experiments on one-dimensional systems of Rydberg atoms\cite{Keesling_2019,  Bernien_2017} has brought back unsolved questions about the nature of the commensurate melting of period-$p$ phases\cite{Ostlund_1981, Huse_Fisher_1982, Huse_Szpilka_1983, Howes_1983, denNijs,Huse_Fisher_1984}. For $p>2$, a floating phase, characterized by incommensurate and algebraic correlations, separates the $\mathbb{Z}_p$-ordered and the disordered phases\cite{haldane_bak,Ostlund_1981,Huse_Fisher_1984,Katsura_2015}. For $p=3$ and $p=4$, the extension of the floating phase is still debated\cite{Selke1982,Duxbury,Huse_Fisher_1984,giudici,prl_chepiga,Whitsitt_2018,chepiga2020kibblezurek, Chepiga_Mila_2021,rader2019floating,NYCKEES2021115365,PhysRevResearch.4.013093}. One open question is the existence of a direct and continuous transition in the chiral universality class that would occur before the floating phase develops\cite{Huse_Fisher_1982,Huse_Fisher_1984}. The quantum version of this problem can be formulated in terms of hard-core bosons associated with domain walls in the commensurate structure of the density wave \cite{Schulz_1982}. The resulting Hamiltonian exhibits a $\mathbb{Z}_p$ symmetry as it contains terms that create and annihilate $p$ adjacent particles. When these perturbations to the free-fermion fixed point are irrelevant, the U(1) charge conservation is restored at the scaling limit in an extended incommensurate Luttinger liquid phase\cite{Fendley_2004, Chepiga_Mila_2021}, equivalent to the floating phase of the 2D classical problem\cite{Huse_Fisher_1984}.

The properties of the Luttinger liquid phase are described by a bosonic conformal field theory\cite{Haldane_1981, Giamarchi}. The correspondence between the operators at the lattice scale and the bosonic fields in the continuum limit can be in principle constructed from selection rules dictated by the symmetry of the lattice Hamiltonian. When the latter exhibits U(1) symmetry, this correspondence is generally given by the standard bosonization mapping\cite{Giamarchi}. For models that do not conserve the number of particles, the latter becomes insufficient to capture the long-distance behavior of the correlations. A similar problem was discussed recently in the case of nonsymmorphic 1D models\cite{Yang_Feb2022, Yang_Aug2022}. We provide in this paper a way to derive a bosonic representation of lattice operators in $\mathbb{Z}_p$-symmetric 1D models. Using the flow equation approach\cite{Wegner}, introduced by Wegner in 1994, a continuous unitary transformation is designed to restore the U(1) symmetry perturbatively in an effective Hamiltonian. The action of these transformations on the lattice operator generates an expansion in terms of the bosonic fields of the Luttinger liquid theory. The bosonization mappings obtained from this procedure are sufficient to capture the long-distance behavior of the correlations. We illustrate this method on two chains of spinless fermions and hard-core bosons where particles are created and annihilated three by three on adjacent sites. Our findings are assessed by density matrix renormalization group (DMRG)\cite{dmrg1,dmrg2,dmrg3,dmrg4} simulations. 

The paper is organized as follows. In Section \ref{section:model}, we discuss the phase diagrams of the hard-core boson and spinless fermion models. Section \ref{section:floweq} provides a brief introduction to the flow equation approach. In Section \ref{section:floweq_fermion}, the flow equation approach is applied to the fermionic model to derive perturbatively a U(1)-symmetric effective Hamiltonian. A modified bosonic representation of the single-fermion operator is then derived and used to calculate correlation functions inside the Luttinger liquid phase. In Section \ref{section:floweq_boson}, the same procedure is applied to the continuum limit of the hard-core boson model, using the generator that diagonalizes the dual sine-Gordon model. The results are summarized in Section \ref{section:Conclusion}.

\section{models and phase diagrams}\label{section:model}
We consider one-dimensional models of hard-core bosons and spinless fermions where particles are created and annihilated three by three on adjacent sites. The two models share the feature of a Luttinger liquid phase that remains stable when the $\mathbb{Z}_3$-symmetric interaction is turned on. In this phase, the low-energy properties of the system are described by the Luttinger liquid Hamiltonian
\begin{equation}\label{LL_theory}
H_{\text{LL}} = \frac{1}{2\pi}\int dx\,vK[\partial_x\theta(x)]^2+\frac{v}{K}[\partial_x\phi(x)]^2,
\end{equation}
where $K$ is the Luttinger parameter and $v$ is the velocity. The fields $\phi$ and $\theta$ are bosonic in nature and satisfy the commutation relation $[\phi(x),\theta(y)]=i\pi\,\text{sign}(y-x)/2$. The correlations decay algebraically with an exponent controlled by $K$ and oscillate with an incommensurate wave vector proportional to the Fermi wave vector $k_F$. We present in this section the arguments for the stability of the Luttinger liquid phase and discuss the nature of the transitions out of it.  
\subsection{Hard-core bosonic model}
The Hamiltonian of the hard-core boson model is given by
\begin{equation}\label{3boson_model}
    H = \sum_{i} -t(b_{i+1}^\dagger b_i+\text{h.c.})-\mu n_i +\lambda(b_i^\dagger b_{i+1}^\dagger b_{i+2}^\dagger + \text{h.c.}),\\
\end{equation}
where $b^\dagger_{i}$ and $b_i$ are respectively the creation and annihilation operators of hard-core bosons at site $i$, and $n_i=b^\dagger_ib_i$ is the density operator. The hard-core constraint amounts to a restriction of the occupation number to $n_i=0$ and $n_i=1$. Accordingly, the operators satisfy the commutation relation $[b_i, b^\dagger_j]=(1-2n_i)\delta_{i,j}$. 
The hard-core boson model was recently introduced as a dual description of the transition to phases with a density-wave order of period 3 in chains of Rydberg atoms~\cite{Whitsitt_2018, Fendley_2004}. Its phase diagram is studied extensively in Ref.~\onlinecite{Chepiga_Mila_2021}. We recall here the main results. Without loss of generality, the hopping amplitude is set to $t=1$. At $\lambda=0$, Eq.\,\eqref{3boson_model} is a free fermion Hamiltonian that can be mapped in the continuum limit to the Luttinger liquid Hamiltonian \eqref{LL_theory} with a Luttinger exponent $K=1$ and a velocity $v = 2\sin(k_F)$. When $\lambda\neq0$, the U(1) symmetry is reduced to a $\mathbb{Z}_3$ symmetry. The stability of the Luttinger liquid phase follows from the scaling analysis of the Hamiltonian in the continuum limit, obtained by applying the bosonization mapping\cite{Giamarchi}. Since the bosonic representation of the hard-core boson operators takes the form $b\sim e^{i\theta}$, the Hamiltonian reduces, at half filling ($\mu=0$) and up to the most relevant term, to
\begin{equation}\label{3boson_bosonization}
H \sim H_{\text{LL}} + \frac{g}{\pi\alpha^2}\int dx \cos(3\theta(x)),
\end{equation}
where $g$ is a dimensionless coupling and $\alpha$ is a real-space cutoff. A standard renormalization group (RG) analysis\cite{Giamarchi} yields the RG equations
\begin{equation}\label{RG_eq}
\begin{split}
\frac{dK(l)}{dl} &= \frac{9}{4}g^2(l),\\
\frac{dg(l)}{dl} &= \Big(2-\frac{9}{4K(l)}\Big)g(l).
\end{split}
\end{equation}
When $K<9/8$, the coupling constant decays exponentially under renormalization such that the Luttinger liquid Hamiltonian is recovered with effective parameters $K^{*}$ and $v^{*}$ that depend on the bare coupling constants in Eq.\,\eqref{3boson_bosonization}. The U(1) symmetry is thus restored at the scaling limit. It manifests itself as a symmetry under translation of the dual field $\theta$. At $K = 9/8$, the system undergoes a Kosterlitz-Thouless (KT) transition\cite{Kosterlitz_Thouless} into a $\mathbb{Z}_3$-ordered phase by pinning the dual field in the minima of the cosine, i.e., at $\theta_n=2\pi n/3$. Upon varying the chemical potential $\mu$ inside the Luttinger liquid phase, a commensurate-incommensurate transition into a disordered phase occurs. This transition is in the Pokrovsky-Talapov (PT) universality class\cite{Pokrovsky_Talapov}, characterized by a dynamical exponent $z=2$ and an incommensurate correlations wave vector that approaches its commensurate value with a singularity proportional to $|\mu-\mu_c|^{1/2}$. The KT and PT lines are expected to meet at a Lifshitz point that would appear before the three-state Potts point\cite{Huse_Fisher_1984}. Between the two points, the commensurate melting is direct and takes place in the chiral universality class. These predictions are confirmed numerically in Ref.~\onlinecite{Chepiga_Mila_2021}.

\subsection{Fermionic model}
The Hamiltonian of the fermionic model is given by
\begin{equation}\label{3fermion_model}
    H = \sum_{i} -t(c_{i+1}^\dagger c_i+\text{h.c.})-\mu n_i +\lambda(c_i^\dagger c_{i+1}^\dagger c_{i+2}^\dagger + \text{h.c.}),\\
\end{equation}
where $c^\dagger_i$ and $c_i$ are the creation and annihilation operator of spinless fermions. It differs from the hard-core boson model \eqref{3boson_model} by a string operator when the Jordan-Wigner transformation is applied. 
\begin{figure}[H]
 \centering
  \includegraphics[scale=0.5]{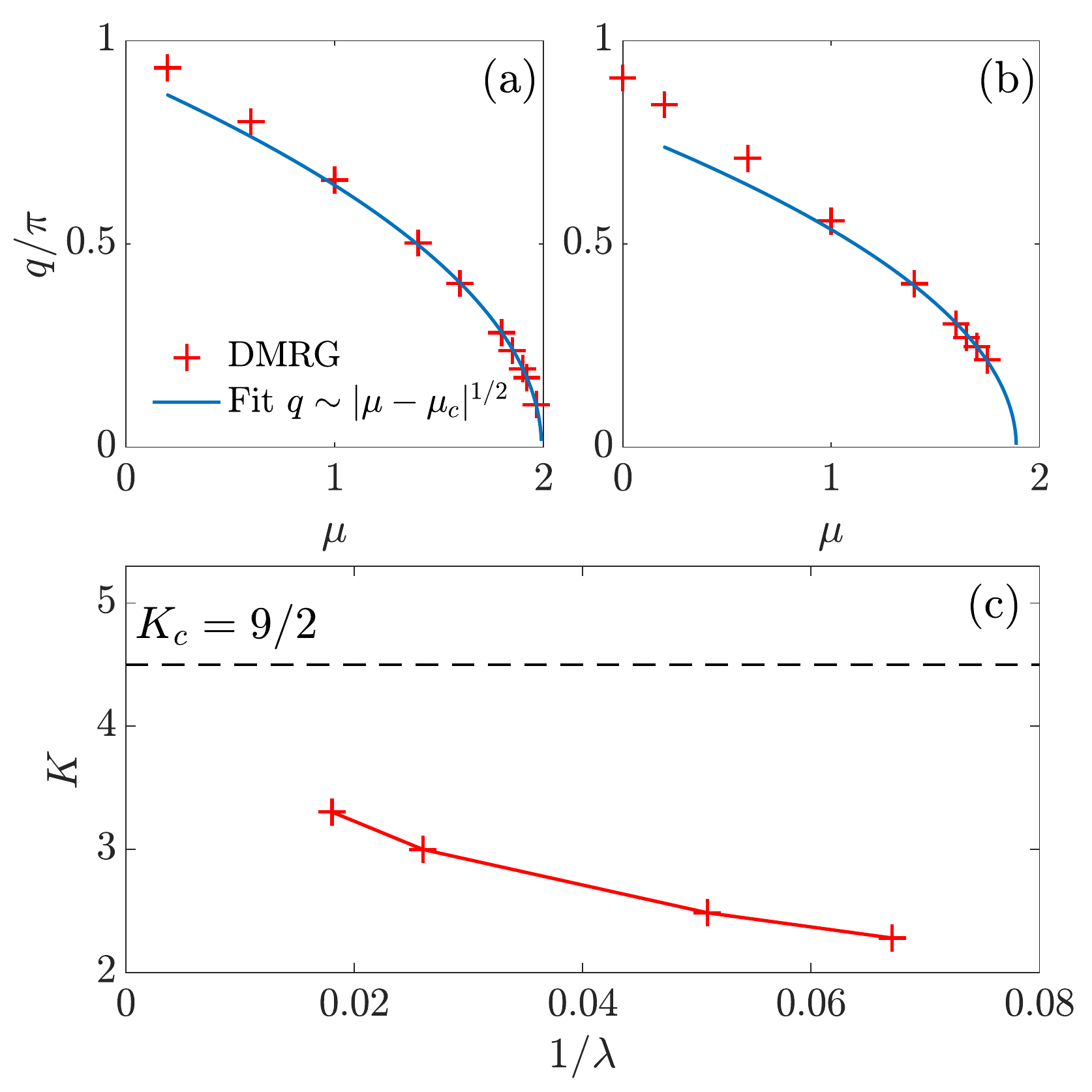}
 \caption{(a)-(b) Scaling of the incommensurate wave vector as a function of the chemical potential close to the PT transition for (a) $\lambda = 0.2$ and (b) $\lambda = 2$. (c) Luttinger parameter $K$ as a function of $\lambda$ at $\mu=0$, extracted numerically from the decay of the two-fermion correlations. This procedure is only valid when $K$ is larger than $\sqrt{3}$ (see Section \ref{subsubsection:fermion_correlation}). The system is simulated with $N=600$ sites.}
 \label{DMRG_fermions}
\end{figure}
Similar to the bosonic case, the stability of the Luttinger liquid phase in Eq.\,\eqref{3fermion_model} can be investigated using the bosonization mapping\cite{Giamarchi}. In terms of the fields $\phi$ and $\theta$, the Hamiltonian reads 
\begin{equation}\label{3fermion_bosonization}
H\sim H_{\text{LL}}+\frac{g}{\pi\alpha^2}\int dx\, \cos(3\phi(x)-3k_Fx)\cos(3\theta(x)),
\end{equation}
where terms that oscillate at $k_F$ are neglected as they vanish under the integration for all fillings of the band inside the Luttinger liquid phase. Eq.\,\eqref{3fermion_bosonization} becomes non-oscillating at $k_F=2\pi/3$. The perturbation to the Luttinger liquid Hamiltonian has a scaling  dimension  $\Delta = 9(K+1/K)/4$ and becomes relevant when $\Delta<2$. This inequality has however no solution and the coupling constant $g$ decays exponentially  under renormalization for all values of $K$.
It should be noted that terms such as $\cos(6\theta)$ and $\cos(6\phi)$ are expected to arise along the RG flow and can lead to a KT transition when $K>9/2$ or $K<2/9$. DMRG computations of $K$ suggest however that it increases slowly enough from $K=1$ for the generated terms to remain irrelevant even at large values of $\lambda$. For instance, the numerical results at $\mu=0$ indicate an extension of the Luttinger liquid phase at least up to $\lambda\sim 50$ (Fig.\ref{DMRG_fermions}). Nevertheless, a KT transition at larger values of $\lambda$ cannot be excluded. As the chemical potential is varied in the Luttinger liquid phase, a PT transition line can be identified numerically using the scaling behavior of the wave vector (Fig.\ref{DMRG_fermions}).

\section{Flow equation approach}\label{section:floweq}
The general idea behind the flow equation approach introduced by Wegner\citep{Wegner} is to apply continuous unitary transformations to the Hamiltonian in order to bring it into a more band-diagonal form. The approach consists in a renormalization scheme where states with large energy differences are first decoupled while smaller energy differences are later suppressed along the flow.

 The formalism of the method is based on the parametrization of a set of unitarily equivalent Hamiltonians $H(l)=U(l)H(0)U^\dagger(l)$. By taking the derivative with respect to $l$, the problem is recast into the differential equation
\begin{equation}
    \frac{dH(l)}{dl} = [\eta(l),H(l)],
\end{equation}
where 
\begin{equation}
\eta(l)=\frac{dU(l)}{dl}U^\dagger(l)
\end{equation}
 is the anti-hermitian generator of the flow. The latter can be chosen appropriately to diagonalize the bare Hamiltonian $H(0)$. The canonical choice proposed by Wegner\cite{Wegner} is given by $\eta(l) = [H_\text{d}(l),H_{\text{od}}(l)]$, where $H_{\text{d}}(l)$ and $H_{\text{od}}(l)$ are the diagonal and the off-diagonal parts of the flowing Hamiltonian $H(l)$. From this definition of the generator, it can be shown\cite{Kehrein} that 
 \begin{equation}
 \frac{d\text{Tr}(H^2_{\text{od}}(l))}{dl}=-2\text{Tr}(\eta^{\dagger}(l)\eta(l))\leq 0,
 \end{equation}
which indicates that the flow gradually brings the Hamiltonian into a more band-diagonal form. A fixed point of the flow is reached when $\eta(l)$ vanishes. The diagonal and off-diagonal parts of the flowing Hamiltonian then commute and the Hamiltonian becomes block-diagonal with respect to the symmetry of the non-interacting part. Thus, the flow equation approach provides a systematic way to design a unitary transformation that recovers the U(1) symmetry in models that do not conserve the number of particles. Its drawback is the proliferation of terms along the flow. Truncations schemes are hence needed to keep the calculations tractable. 

Once the diagonal Hamiltonian is obtained from the flow equation procedure, the change of basis associated with $\eta(l)$ can be applied to the operators. Given an operator $O$, its transformation along the flow is dictated by the flow equation
\begin{equation}
\frac{dO(l)}{dl} = [\eta(l),O(l)].
\end{equation}
The expectation value in the ground state of the bare Hamiltonian can be evaluated using the relation
\begin{equation}
\bra{\psi_{\text{gs}}}O\ket{\psi_{\text{gs}}} = \bra{\psi_{\text{gs}}(\infty)}O(\infty)\ket{\psi_{\text{gs}}(\infty)},
\end{equation}
where $\ket{\psi_{\text{gs}}(\infty)} = U^\dagger(\infty)\ket{\psi_{\text{gs}}}$ is the ground state of the diagonal Hamiltonian $H(\infty)$.

In the following sections, we apply the flow equation procedure to the models in Eqs.\,\eqref{3boson_model},\eqref{3fermion_model}. The transformed bosonic representations of hard-core bosonic and fermionic operators is derived from the generator of the flow that restores the U(1) symmetry in the Hamiltonians.
\section{Flow equation approach to\\the fermionic model}\label{section:floweq_fermion}
We derive  in this section a U(1)-symmetric effective Hamiltonian that describes the low-energy properties of the fermionic Hamiltonian \eqref{3fermion_model}.  
\subsection{Flow of the Hamiltonian}
We proceed by writing the Hamiltonian in the Fourier basis in order to separate it into a diagonal part $H_0$ and an interaction part $\lambda H_3$. We have
\begin{equation}
    H = \sum_{k}\xi_k c_{k}^\dagger c_{k}+\frac{\lambda}{3!\sqrt{N}}\sum_{k,q}B_{k,q}(c_{k}^\dagger c_{q}^\dagger c_{-k-q}^\dagger - \text{h.c.}),\\
\end{equation}
where 
\begin{equation}
	\begin{split}
	\xi_k  &= -2\cos(k)-\mu\\
    B_{k,q} &= 2i[\sin(2k+q)-\sin(2q+k)-\sin(k-q)].
    \end{split}
\end{equation}
Along the flow, other interaction terms that are not initially present are generated. They are incorporated in the following ansatz for the flowing Hamiltonian $H(l)$:
\begin{equation}\label{ansatz}
    H(l) =  H_0(l) + \lambda H_{3}(l)+\lambda^2 H_{U}(l),
\end{equation}
where
\begin{equation}\label{H_U}
 H_{U}(l) = \frac{1}{N}\sum_{k,q,p}U_{k,q,p}(l)c^\dagger_{k+p}c^\dagger_{q-p}c_{q}c_{k}.
\end{equation}
The normal-ordering of terms in Eq.\eqref{ansatz} with respect to the Fermi sea of the diagonal Hamiltonian $H_0(l)$ is implicit. It is systematically carried out during the calculation to truncate the generated terms in a controlled manner. We note that three-body interaction and six-fermion terms also arise along the flow. They are neglected in the ansatz \eqref{ansatz} due to their large scaling dimension. By taking the generator as $\eta(l) := [H_0(l),\lambda H_3(l)]$, which reads
\begin{equation}\label{generator}
 \eta(l) = \frac{\lambda}{3!\sqrt{N}}\sum_{k,q}B_{k,q}(l)\alpha_{k,q}(l)(c^\dagger_{k}c^\dagger_{q}c^\dagger_{-k-q}+\text{h.c.}),
\end{equation}
with $\alpha_{k,q}=\xi_{k}+\xi_{q}+\xi_{-k-q}$, we obtain a set of flow equations:
\begin{equation}\label{Flow_Eq}
	\begin{split}
	\frac{dB_{k,q}(l)}{dl} &= -\alpha^2_{k,q}(l)B_{k,q}(l),\\
	\frac{dU_{k,q,p}(l)}{dl} &= -\frac{1}{2}[\alpha_{k+p,q-p}(l)+\alpha_{k,q}(l)]B_{k,q}(l)B_{k+p,q-p}(l)\\
	&\hspace{2cm}\times [2\Theta(k_F-|k+q|)-1]\\
	\frac{d\xi_{k}(l)}{dl} &= -\frac{\lambda^2}{N}\Big[\sum_{|q|<k_F}\alpha_{k,q}(l)B^2_{k,q}(l)\Theta(k_F-|k+q|)\\
	&+\sum_{|q|>k_F}\alpha_{k,q}(l)B^2_{k,q}(l)[1-\Theta(k_F-|k+q|)]\Big]
	\end{split}
\end{equation}
with the initial conditions $B_{k,q}(0)=B_{k,q}$, $\xi_{k}(0)=\xi_k$, and $U_{k,q,p}(0)=0$. We have omitted for brevity the $l$-dependence in the bare Hamiltonian, i.e. $H\equiv H(0)$. The derivation of the flow equations is detailed in Appendix \ref{appendix:Derivation_floweq}. To the leading order in $\lambda$, it is sufficient to take the bare value $\alpha_{k,q}$ in the flow equation of $B_{k,q}(l)$. The solution shows that the three-site term decays to zero along the flow as $B_{k,q}(l)=B_{k,q}e^{-\alpha^2_{k,q}l}$. On the other hand, the generated terms take finite values at $l=\infty$ given by
\begin{equation}\label{Flow_solution}
	\begin{split}
	U_{k,q,p}(\infty) &= -\frac{1}{2}\frac{\alpha_{k+p,q-p}+\alpha_{k,q}}{\alpha^2_{k,q}+\alpha^2_{k+p,q-p}}B_{k,q}B_{k+p,q-p}\\
	&\hspace{2cm}\times [2\Theta(k_F-|k+q|)-1]\\
	\xi_k(\infty) &= \xi_k-\frac{\lambda^2}{2N}\Big[\sum_{|q|<k_F}\frac{B^2_{k,q}}{\alpha_{k,q}}\Theta(k_F-|k+q|)\\&\hspace{1cm}+\sum_{|q|>k_F} \frac{B_{k,q}^2}{\alpha_{k,q}}[1-\Theta(k_F-|k+q|)]\Big]
	\end{split}
\end{equation}
By construction, the two-body interaction matrix $U_{k,q,p}$ is symmetrized with respect to the permutations of fermion operators in Eq.\,\eqref{H_U}. The divergence as $\alpha_{k,q}$ goes to zero in the renormalized quantities \eqref{Flow_solution} is an artifact of the limit $l\to\infty$. In fact, the contribution of the flow to the dispersion vanishes when $\alpha_{k,q}=0$. Finally, we note that Wegner's prescription is relaxed by taking in the definition \eqref{generator} of $\eta(l)$ only the three-site term and not all the interacting part of $H(l)$. As a result, the approach becomes perturbative in $\lambda$. Terms that are generated along the flow have a second-order dependence in $\lambda$ and can be eliminated by including them in a redefinition of the generator. This induces higher-order corrections to the flow equations \eqref{Flow_Eq}. Thus, the flow equation procedure enables to push the $\lambda$-dependence of terms that break the U(1) symmetry to higher orders, thereby restoring the symmetry perturbatively.

\subsection{Bosonization of the effective Hamiltonian}\label{section:bosonization}
We investigate in this section  the low-energy properties of the Hamiltonian \eqref{3fermion_model} starting from the effective Hamiltonian $H_{\text{eff}}=H_0(\infty)+\lambda^2H_{U}(\infty)$. After taking the long-range limit in the two-body interaction term, only particle-hole excitations within a momentum range $\Lambda$ around the Fermi points $\pm k_F$ are retained. The renormalized dispersion relation can then be linearized around the Fermi point and the fermion operators separated into right and left modes: 
\begin{equation}\label{c_k_separation}
c_{k} = \Theta(\Lambda-|k-k_F|)c_{R,k}+\Theta(\Lambda-|k+k_F|)c_{L,k},
\end{equation}
where $\Theta$ is the Heaviside function. After constructing fermionic fields from these modes, i.e.,
\begin{equation}\label{fermion_field}
c(x) = \frac{1}{\sqrt{N}}\sum_{k}e^{ikx}c_{k}=:c_R(x)+c_L(x),
\end{equation}
we obtain in the continuum limit a Hamiltonian that describes the low-lying states of $H_{\text{eff}}$. It is given by
\begin{equation}\label{Tomonaga-Luttinger}
\begin{split}
H_{\text{eff}} =& -i\Tilde{v}_F\int dx\,\big[c^\dagger_R(x)\partial_xc_R(x)-c^\dagger_L(x)\partial_xc_L(x)\big]\\
&+ 4\lambda^2 g_2 \int dx\,\rho_R(x)\rho_L(x),
\end{split}
\end{equation}
where $\Tilde{v}_F=\partial\xi_k(\infty)/\partial k|_{k=k_F}$ is the renormalized Fermi velocity, $\rho_{R}(x)$ and $\rho_L(x)$ are respectively the density operators of the right and left branches, and $g_2=U_{k_F,-k_F,0}$ is the forward scattering matrix element. Due to the symmetry of the two-body interaction matrix, $g_4$ scattering processes that couple fermions at the same branch vanish. Finally, the factor 4 accounts for the two possible $g_2$ processes and the two backscattering processes, i.e., $g_1=-U_{k_F,-k_F,2k_F}$, which for spinless fermions coincide with $g_2$ processes. We note that since $g_2>0$, the interaction is attractive, which indicates that the three-site term in the bare Hamiltonian favors the occupation of three adjacent sites.

We now apply the bosonization mapping between the fermionic fields $c_R,c_L$ and the bosonic fields $\phi, \theta$. It is given by\,\cite{Giamarchi}
\begin{equation}\label{Bosonization}
c_r(x) = \frac{F_r}{\sqrt{2\pi\alpha}}e^{irk_Fx}e^{-i[r\phi(x)-\theta(x)]},
\end{equation}
where $r=1$ for $r=R$ and $r=-1$ for $r=L$. Here, $\alpha\sim 1/\Lambda$ is a short-distance cut-off and $F_r$ are unitary operators called Klein factors. They follow the commutation relations $\{F_r,F^\dagger_{r'}\} = 2\delta_{r,r'}$ and ensure the anti-commutation of fermions from different species. Using Eq.\,\eqref{Bosonization}, the Hamiltonian \eqref{Tomonaga-Luttinger} can be reduced to the Luttinger liquid Hamiltonian \eqref{LL_theory} with a renormalized velocity $u$ and a Luttinger parameter, given to the second order of $\lambda$ by
\begin{equation}\label{K_prediction}
K = 1+\frac{4}{\pi}\sin(k_F)\sin^2\Big(\frac{k_F}{2}\Big)\lambda^2.
\end{equation}
The derivation of Eq.\,\eqref{K_prediction} is detailed in Appendix \ref{appendix:Derivation_K}. Here, $k_F$ is defined by the filling of the renormalized band $\xi_k(\infty)$ and differs from its value at the non-interacting limit $\lambda=0$. The relation between $k_F$ in Eq.\,\eqref{K_prediction} and the bare chemical potential can be obtained by setting $\xi_{k_F}(\infty)=0$. This leads to the auto-coherent equation
\begin{equation}\label{bare_chempot}
\mu = -2\cos(k_F)-\lambda^2I(k_F,\mu),
\end{equation}  
where $I$ is the sum over the $q$ modes in Eq.\,\eqref{Flow_solution}.

\begin{figure}[H]
\begin{center}
\centering
 \includegraphics[scale=0.55]{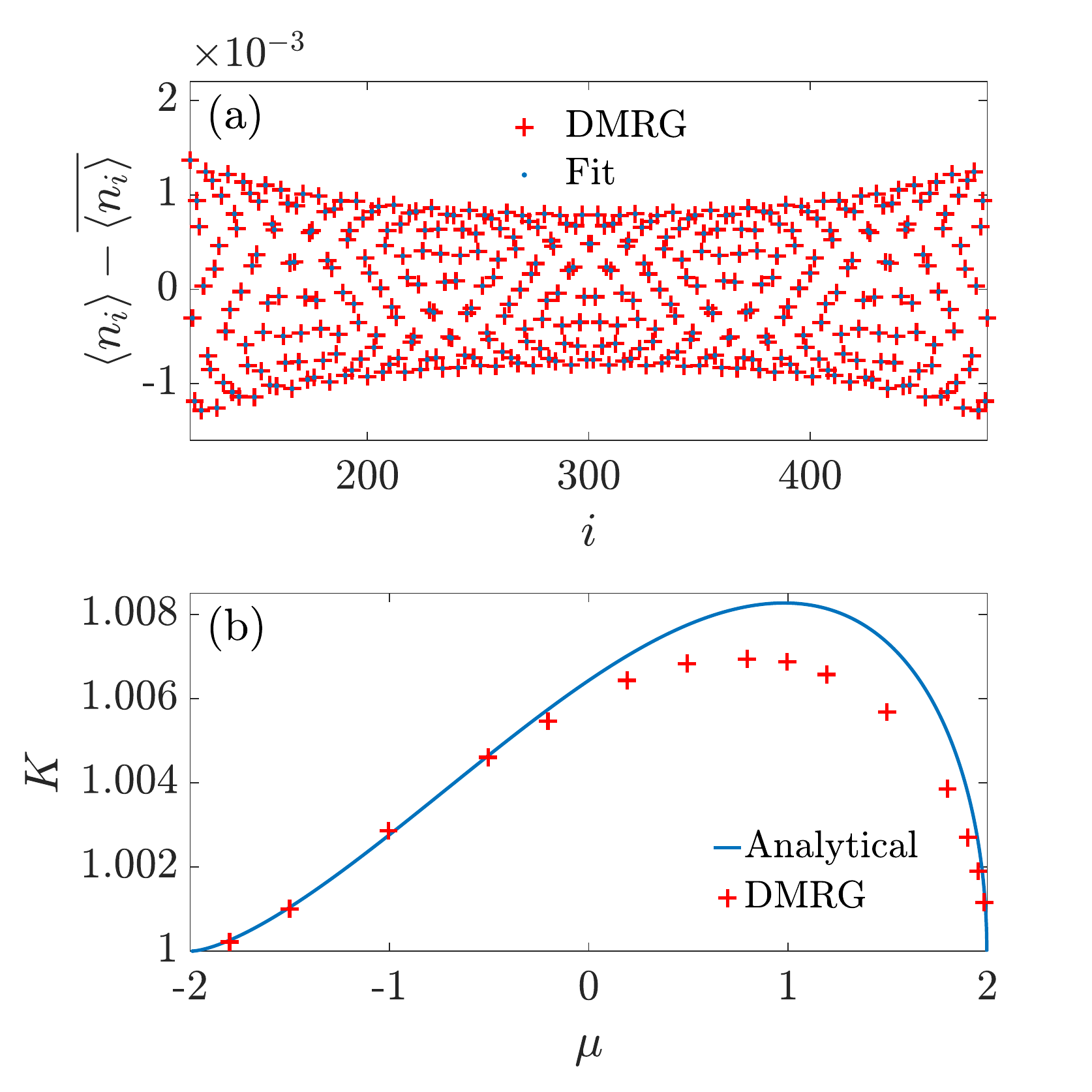}
 \caption{(a) Example of the local density profile at $\lambda=0.1$ and $\mu=1.2$. (b) Analytical calculation and DMRG simulations of the Luttinger parameter $K$ at $\lambda=0.1$ as a function of the bare chemical potential $\mu$, obtained by solving Eq.\,\eqref{bare_chempot}. }
  \label{K_DMRG}
  \end{center}
\end{figure}

The behavior of the Luttinger parameter is compared to DMRG results, along a small $\lambda$ cut, where the perturbative calculations still hold (Fig.\ref{K_DMRG}). $K$ is obtained numerically by fitting the profile of the local density, which for open boundary conditions exhibits Friedel oscillations. According to conformal field theory, we have \cite{White_Affelck_2002, Fabrizio_Gogolin_1995}
\begin{equation}
\expval{n_j}\propto \frac{\cos(2k_Fj+\beta)}{[(N/\pi)\sin(\pi j/N)]^K},
\end{equation} 
with a phase shift $\beta$.
\noindent The DMRG results for $K$ follow the form described by Eq.\,\eqref{K_prediction}. The absence of a reflection symmetry with respect to $\mu$ is due to the breaking of particle-hole symmetry by the three-site interaction in Eq.\,\eqref{3fermion_model}.
The maximal deviation between the analytical and the numerical results is of order $10^{-3}$. Its origin is twofold. First, it should be noted that by considering only scattering processes with momentum transfer $p\sim0,2k_F$ in the effective Hamiltonian \eqref{Tomonaga-Luttinger}, we have neglected terms that oscillate with a wave vector  $2k_F$. These terms contribute to the renormalization of $K$ at the $4^{\text{th}}$ order of $\lambda$, mainly at half-filling. Secondly, three-body and six-fermion interaction terms are discarded in the early stages of the calculations. Including them in the generator of the flow also provides corrections of order $4$ in $\lambda$ to the two-body interaction matrix.

\begin{figure}[H]
\centering
 \includegraphics[scale=0.5]{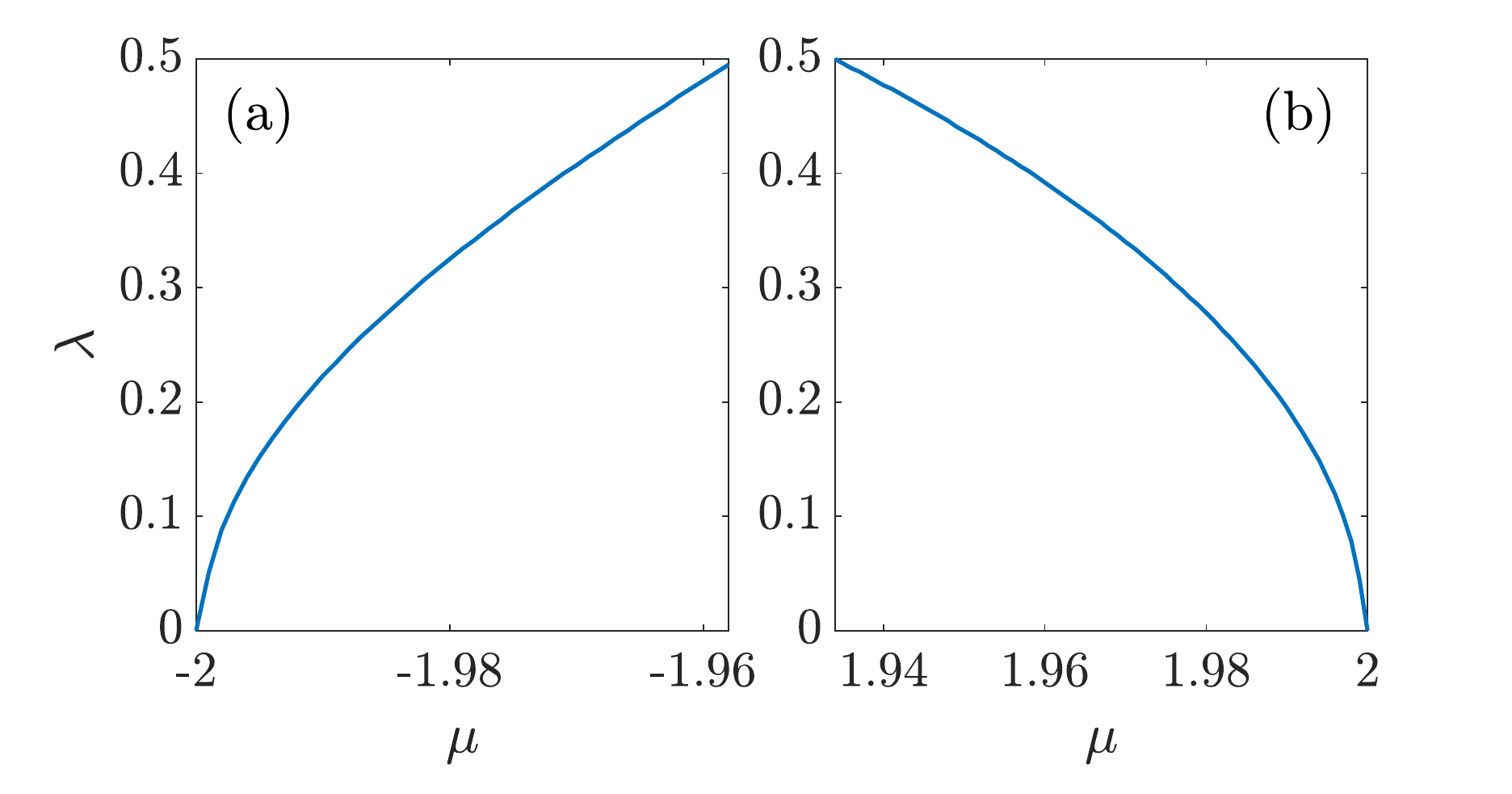}
 \caption{(a)-(b) Theoretical prediction for the Pokrovsky-Talapov transition line at (a) $k_F=0$ and (b) $k_F=\pi$, as a function of $\lambda$.}
 \label{PT_line}
\end{figure}

When the edges of the band $\xi_k(\infty)$ are crossed, the system undergoes a PT transition as the density of fermions respectively vanishes or saturates. For small $\lambda$, the shape of the transition line can be extracted from the solution of Eq.\,\eqref{bare_chempot} at $k_F=0,\pi$ (Fig.\ref{PT_line}). As the transition is approached from the Luttinger liquid phase, the two-body interaction matrix vanishes, i.e., $U_{k_F,-k_F,p}=0$ for all $p$ and the Luttinger parameter tends, up to the second order in $\lambda$, towards its non-interacting value $K=1$. This result indicates that although the particles are not conserved in the bare Hamiltonian, they still behave as free fermions in the low density limit.
\subsection{Transformed Fermionic operator}\label{section:transformed_fermion}
In this section, a modified bosonic representation of the single fermion operator is derived from the flow to the effective Hamiltonian \eqref{Tomonaga-Luttinger}. Consequences of these modifications on the correlation functions of bare operators are discussed.
\subsubsection{Flow of single fermion operators}
In order to evaluate observables in the ground state of $H_{\text{eff}}$, the operators need also to be transformed. In particular, the form of the single fermion operator under the transformation $U(\infty)$ that block-diagonalizes the Hamiltonian can be obtained by solving the flow equation 
\begin{equation}\label{flow_eq_c}
\frac{dc_k(l)}{dl} = [\eta(l),c_k(l)].
\end{equation}
Similar to the flow of the Hamiltonian, a closed form of the solution to Eq.\,\eqref{flow_eq_c} is not tractable and truncations should be carried out. We take here the ansatz
\begin{equation}\label{c_k_transformed}
\begin{split}
c_{k}(l) = c_{k}&+\frac{\lambda}{\sqrt{N}}\sum_{q}\gamma_{k,q}(l)c^\dagger_{q}c^\dagger_{-k-q}\\
&-\frac{2\lambda}{3\sqrt{N}}\sum_{q,p}\gamma_{q,p}(l)\big[c^\dagger_{q}c^\dagger_{p}c^\dagger_{-p-q}-c_{q}c_{p}c_{-p-q}\big]c_k.
\end{split}
\end{equation}
It will later be argued that higher-order terms in $\lambda$ do not bring any qualitative change to the behavior of the correlation functions. The flow equation reads
\begin{equation}
\frac{d\gamma_{k,q}(l)}{dl} = -\frac{1}{2}B_{k,q}(l)\alpha_{k,q}(l),
\end{equation}
with the initial condition $\gamma_{k,q}(0) = 0$. The solution is given by $\gamma_{k,q}(\infty)=-B_{k,q}/2\alpha_{k,q}$. Since the U(1) symmetry is recovered in the rotated basis, the standard bosonization mapping \eqref{Bosonization} can be applied. We start by constructing a fermionic field  $\Tilde{c}(x)$ from the transformed modes $c_{k}(\infty)$ in the same fashion as in Eq.\,\eqref{fermion_field}. Given that the mode $k$ decouples from the other modes in the last term of Eq.\,\eqref{c_k_transformed}, the Fourier transform yields a local operator that will at most lead to a renormalization of the numerical prefactors in correlation function. Hence, we only consider
\begin{equation}\label{psi(x)_transformed}
\begin{split}
\Tilde{c}(x) \sim\,c(x)+\frac{\lambda}{N}\int dydz\,\Gamma(z,z-y)c^\dagger(x+y)c^\dagger(x+z),\\
\end{split}
\end{equation}
where
\begin{equation}
\Gamma(x,y) = \frac{1}{N}\sum_{k,q}e^{-ikx}e^{-iqy}\gamma_{k,q}(\infty).
\end{equation}
The most relevant operators in the bosonic representation of Eq.\,\eqref{psi(x)_transformed} are extracted by carrying out operator product expansions (OPE) in the product of vertex operators. We have
\begin{equation}\label{OPE}
    \begin{split}
    c^\dagger(x+y)c^\dagger(x+z) &\sim\frac{1}{\pi\alpha}\sin(k_F(z-y))e^{-2i\theta(x)}.
    \end{split}
\end{equation}
By inserting this expression back into Eq.\,\eqref{psi(x)_transformed} and carrying out the double integration, we obtain
\begin{equation}\label{c_bos}
\begin{split}
    \Tilde{c}(x) \sim\,&\frac{e^{ik_Fx}}{\sqrt{2\pi\alpha}}e^{-i[\phi(x)-\theta(x)]} + \frac{e^{-ik_Fx}}{\sqrt{2\pi\alpha}}e^{i[\phi(x)+\theta(x)]}\\
     &+ \frac{4\lambda}{\pi\alpha} \sin(k_F) e^{-2i\theta(x)}.\\
    \end{split}
\end{equation}
From the structure of the generator $\eta$, the form of higher-order terms in the transformed operator can be guessed. We will only consider operators of order $\lambda^2$ as these can contribute to the second-order expansion of the correlations by combining with the $\text{zero}^{\text{th}}$-order part of Eq.\,\eqref{c_k_transformed}. Given an operator $O(c,c^\dagger)$ generated at the second order in $\lambda$ , it transforms under U(1) rotations as $O(e^{i\alpha}c, e^{-i\alpha}c^\dagger) = e^{in\alpha}O(c,c^\dagger)$, where $n$ can only take the values $1,7,-5$. Since the U(1) rotations act on the bosonic fields as $\phi\to\phi$ and $\theta\to\theta+\alpha$, this indicates that operators with a scaling dimension smaller than those in Eq.\,\eqref{c_bos} cannot be generated. The terms that transform as a single-fermion operator will merely add a $\lambda^2$-dependence to the prefactor of the corresponding vertex operators. Thus, the transformed fermion operator truncated to the second order in $\lambda$ takes the form
\begin{equation}\label{c_bos_effective}
\begin{split}
    \Tilde{c}(x) \sim\,&C_1e^{ik_Fx}e^{-i[\phi(x)-\theta(x)]}\\ &+ C_1e^{-ik_Fx}e^{i[\phi(x)+\theta(x)]}
     + C_2e^{-2i\theta(x)},\\
    \end{split}
\end{equation}
with $C_1 \sim 1+O(\lambda^2)$ and $C_2\sim O(\lambda)$.
\subsubsection{Fermionic correlation functions}\label{subsubsection:fermion_correlation}
Using the effective bosonic representation of the transformed fermion operator \eqref{c_bos_effective}, we can now compute correlation functions in the Luttinger liquid phase. Consider the point-split product of bare-fermion operator
\begin{equation}\label{F_p}
F_p(x) = \lim_{\Delta\to 0}\prod_{n=0}^{p-1} c(x+n\Delta).
\end{equation}
The correlations in the ground state of the bare Hamiltonian can be evaluated in the Luttinger liquid ground state of $H_{\text{eff}}$ using the relation
\begin{equation}\label{corr_relation}
\expval{F_p(x)^\dagger F_p(y)}_{\text{GS}} = \expval{\Tilde{F}^\dagger_p(x)\Tilde{F}_p(y)}_{\text{LL}},
\end{equation}
where $\Tilde{F}_p(x) = U(\infty)F_p(x)U^\dagger(\infty)$ is obtained by replacing $c$ with the transformed field $\Tilde{c}$ in Eq.\,\eqref{F_p}. The correlations of $p$ fermions are deduced from the well-known result\,\citep{Giamarchi} for the correlation function of vertex operators in the Luttinger liquid Hamiltonian \eqref{LL_theory}:
\begin{equation}\label{correlation_LL}
\expval{e^{i[n\phi(x)+m\theta(x)]}e^{-i[n\phi(y)+m\theta(y)]}}_{\text{LL}} \sim \frac{1}{|x-y|^{\frac{n^2K}{2}+\frac{m^2}{2K}}}.
\end{equation}
Accordingly, Eqs.\,\eqref{c_bos_effective} and \eqref{corr_relation} yield the one-fermion correlations
\begin{equation}\label{correlation_F1}
\begin{split}
\expval{F_1^\dagger(x)F_1(y)}_\text{GS} \sim &\,\,2(C_1)^2\frac{\cos(k_Fr)}{r^{\frac{1}{2K}+\frac{K}{2}}}+(C_2)^2\frac{1}{r^{\frac{2}{K}}},
\end{split}
\end{equation}
where $r=|x-y|$. Similarly, the two-fermion correlations can be obtained from the bosonic representation of $F_2$. We have
\begin{equation}\label{F_2}
\begin{split}
\Tilde{F}_2(x) \sim\,\, &(C_1)^2e^{2i\theta(x)}+C_1C_2e^{ik_Fx}e^{-i[\phi(x)+\theta(x)]}\\
&+C_1C_2e^{-ik_Fx}e^{i[\phi(x)-\theta(x)]},
\end{split}
\end{equation}
which leads to
\begin{equation}\label{correlation_F2}
\begin{split}
\expval{F_2^\dagger(x)F_2(y)}_\text{GS} \sim &\,\,(C_1)^4\frac{1}{r^{\frac{2}{K}}}+2(C_1C_2)^2\frac{\cos(k_Fr)}{r^{\frac{1}{2K}+\frac{K}{2}}}.
\end{split}
\end{equation}
\begin{figure}[H]
\centering
  \includegraphics[width=0.95\textwidth]{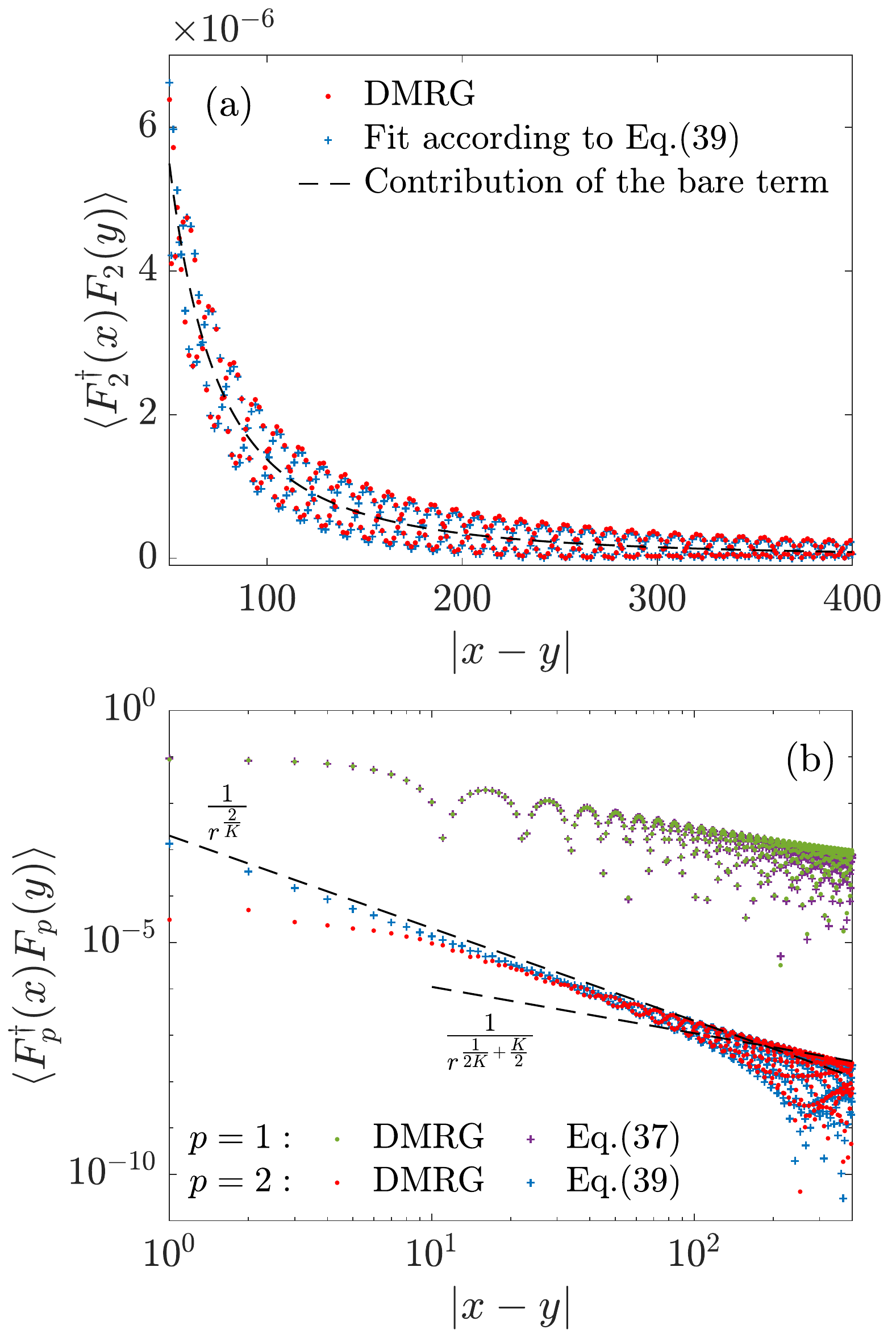}
 \caption{One fermion and two-fermion correlation functions in the ground state of the bare Hamiltonian \eqref{3fermion_model} at $\mu=1.86$ and $\lambda=3$. The dashed line are provided as a guide to the eye. The DMRG calculations are performed on a system of size $N=2100$. The Luttinger parameter $K$ is extracted from Friedel oscillations as discussed in section \ref{section:bosonization} and the coefficients in Eqs.\,\eqref{correlation_F1} and \eqref{correlation_F2} are obtained from a least square fit.}
 \label{DMRG_correlation}
\end{figure}
We note that a single-fermion operator appears in the bosonic representation of $F_2$. It is an example of a term that is not initially present in the standard bosonization mapping but is generated by the $\mathbb{Z}_3$-symmetric interaction along the flow. Its consequence is a crossover between two power-laws in the two-fermion correlations. At short distance, the correlations decay with the standard exponent $2/K$ of the two-fermion operator. At large distance, a single-fermion part with an exponent $\frac{K}{2}+\frac{1}{2K}$, oscillating with a wave vector $k_F$, takes over the correlations when $K<\sqrt{3}$. The crossover takes place at length scale $l\sim 1/\lambda^{4K/(3-K^2)}$. Numerically, it can be observed near the PT transition line, where the Luttinger parameter tends to $K=1$ while $\lambda$ remains small enough for $l$ to remain smaller than the system size. The numerical results for the correlations (Fig.\ref{DMRG_correlation}), fitted with Eqs.\,\eqref{correlation_F1},\eqref{correlation_F2} by the least-square method, are in concordance with the analytical calculations. Finally, it should be noted that a two-fermion operator is also generated in the bosonic representation of the single fermion operator (see Eq.\,\eqref{c_bos_effective}). Its effect on the correlation remains however small compared to the single-fermion part.

\section{Flow equation approach to\\the hard-core bosonic model}\label{section:floweq_boson}
We turn now to the hard-core boson model \eqref{3boson_model}. The commutation relation of the hard-core boson operator renders the calculation of the flow on the lattice difficult. It is more convenient to consider the Hamiltonian in the continuum limit, where it reduces to the dual sine-Gordon model:
\begin{equation}\label{dual_sG}
H = \frac{v_F}{2\pi}\int dx\,[\partial_x\theta(x)]^2+ [\partial_x\phi(x)]^2 +\frac{g}{\pi\alpha^2}\int dx \cos(\beta\theta(x)),
\end{equation}
with $\beta = 3/\sqrt{K}$. Eq.\,\eqref{dual_sG} is obtained from the bosonization mapping \eqref{3boson_bosonization} by absorbing the Luttinger parameter $K$ into a redefinition of the bosonic fields, i.e., $\phi\to \phi\sqrt{K}$ and $\theta\to \theta/\sqrt{K}$. The duality transformation $\phi\leftrightarrow\theta$ and $K\to 1/K$ recovers the sine-Gordon Hamiltonian, which is studied with the flow equation approach in Refs.\onlinecite{Kehrein_1999, Kehrein_2001}. We review here the main result of this work and establish the duality correspondence with Eq.\,\eqref{dual_sG} to obtain a low-energy effective Hamiltonian. 
\subsection{Flow of the sine-Gordon Hamiltonian}
We introduce the vertex operators $V_r(\beta, x) =\,:e^{i\beta[r\phi(x)-\theta(x)]}:$, where $r=R,L$ denotes the left and right species. The relation between our notations and those used in the reference can be found in Appendix \ref{appendix:Bosonization_dic}. The columns refer to the normal ordering with respect to the ground state of the non-interacting part in Eq.\,\eqref{dual_sG}. After performing the duality transformation, the interaction part of the Hamiltonian \eqref{dual_sG} reads
\begin{equation}\label{sG_int}
H_3=\frac{g}{2\pi\alpha^2}\Big(\frac{2\pi\alpha}{L}\Big)^{\frac{\Tilde{\beta}^2}{4}}\hspace{-0.1cm}\int \hspace{-0.1cm}dx \big[V_R(\Tilde{\beta}/2,x)V_L(-\Tilde{\beta}/2,x)+\text{h.c.}\big],
\end{equation}
with $L$ the total length of the chain and $\Tilde{\beta}(K)=\beta(1/K)$. Combined with the non-interacting part, Eq.\,\eqref{sG_int} is the starting point of calculations of the flow equations carried out in Ref.\onlinecite{Kehrein_1999,Kehrein_2001}. The Hamiltonian is diagonalized by a generator $\eta(l)=\eta^{(1)}(l)+\eta^{(2)}(l)$, where
\begin{equation}\label{generator_sG}
\begin{split}
\eta^{(1)}(l) &= -2iv_F\int dxdy\frac{\partial u(y,l)}{\partial y}\\&\hspace{1.5cm}\times\big[V_R(\Tilde{\beta}/2,x)V_L(-\Tilde{\beta}/2,x-y)+\text{h.c.}\big],\\
\end{split}
\end{equation}
and $u(x,l)$ is obtained from the Fourier transform of $u(k,l) = \frac{g(l)}{4\pi^2\alpha^2}\Big(\frac{2\pi\alpha}{L}\Big)^{\frac{\Tilde{\beta}^2}{4}}e^{-4v_F^2k^2l}$. Here, $g(l)$ is a running coupling that flows to zero in the weak-coupling regime, i.e., inside the Luttinger liquid phase. It is initially given by the bare coupling constant in Eq.\,\eqref{dual_sG}. $\eta^{(2)}(l)$ generates the flow of the parameter $\beta$. Its expression can be found in Ref.\,\onlinecite{Kehrein_2001}. At the end of the flow, an effective Hamiltonian $H(\infty)=H_0+H_{\text{d}}(\infty)$ is obtained, where $H_0$ denotes the non-interacting part in the bare Hamiltonian and 
\begin{equation}\label{H_d}
\begin{split}
H_{\text{d}}(\infty) = \sum_{k>0}\omega_k(\infty)\Big[\Tilde{P}_R(-k)\Tilde{P}^\dagger_R(-k)+\Tilde{P}^\dagger_R(k)\Tilde{P}_R(k)\\
+\Tilde{P}_L(k)\Tilde{P}^\dagger_L(k)+\Tilde{P}^\dagger_L(-k)\Tilde{P}_L(-k)\Big],
\end{split}
\end{equation}
with $\omega_k(\infty) = -v_Fg^2\frac{\cos(\pi\Tilde{\beta}^2/4)}{2\Gamma^2(\Tilde{\beta}^2/4)}k|\alpha k|^{(\Tilde{\beta}^2-8)/2}$. $\Tilde{P}_R(k)$ and $\Tilde{P}_L(k)$ are soliton and antisoliton creation and annihilation operator defined as the Fourier transform of vertex operators (see Appendix \ref{appendix:Bosonization_dic}). The effective Hamiltonian of the dual sine-Gordon Hamiltonian \eqref{dual_sG} is deduced from the dual transformation of Eq.\,\eqref{H_d}. The latter acts on the soliton and antisoliton operators as $\Tilde{P}_R(k)\to P^\dagger_R(-k)$ and $\Tilde{P}_L(k)\to P_L(k)$, where $P_r(k)$ is obtained from $\Tilde{P}_r(k)$ by replacing $\Tilde{\beta}$ with $\beta$. Moreover, the flow equations for $g$ and $K$ in Eq.\,\eqref{dual_sG} can be deduced (see Appendix \ref{appendix:Derivation_floweq}):
\begin{equation}\label{floweq_dualsG}
\begin{split}
\frac{dK(l_{\text{RG}})}{dl_{\text{RG}}}&=\frac{9}{4\Gamma(\frac{9}{4K(l_{RG})}-1)}g^2(l_{\text{RG}}),\\
\frac{dg(l_{\text{RG}})}{dl_{\text{RG}}}&=\Big(2-\frac{9}{4K(l_{RG})}\Big)g(l_{\text{RG}}),
\end{split}
\end{equation}
where $l_{\text{RG}}$ is the parameter of the RG flow. It is related to the parameter of the flow equation approach $l$ by $l_{\text{RG}} = \frac{1}{2}\ln(\frac{32l}{\alpha^2})$. The RG equations, derived in Eq.\,\eqref{RG_eq}, are recovered by an expansion of the Gamma function around the critical value $K_c=9/8$. Finally, we note that since the soliton and antisoliton operators transform under U(1) rotations as $P_r(k)\to e^{i\alpha}P_r(k)$, the U(1) symmetry is restored in the effective Hamiltonian. 
\subsection{Transformed hard-core boson operator}
From the Jordan-Wigner transformation and the bosonization mapping \eqref{Bosonization}, a bosonic representation of the hard-core boson can be derived\,\cite{Giamarchi}. It is given by
\begin{equation}\label{hard-core_bosonization}
b(x) = \frac{e^{i\theta(x)}}{\sqrt{2\pi\alpha}}\big[1+\cos(2\phi(x)-2k_Fx)\big].
\end{equation} 
We calculate in this section the transformation of Eq.\,\eqref{hard-core_bosonization} along the flow that diagonalizes the dual sine-Gordon model.
\subsubsection{Flow of the hard-core boson operator}
The flow of the hard-core boson operator is evaluated using the generator of the dual sine-Gordon Hamiltonian. The latter is given by the dual of Eq.\,\eqref{generator_sG}, i.e.,
\begin{equation}\label{generator_dualsG}
\begin{split}
\eta_{\text{dual}}^{(1)}(l) &= -2iv_F\int dxdy\frac{\partial u(y,l)}{\partial y}\\&\hspace{1.5cm}\times\big[V_R(\beta/2,x)V_L(\beta/2,x-y)+\text{h.c.}\big].\\
\end{split}
\end{equation}
We take the following ansatz for the flowing hard-core boson operator, truncated to the most relevant terms:
\begin{equation}\label{hard-core_ansatz}
\begin{split}
b(x,l) &= [C_1(l)+C_2(l)\cos(2\phi(x)-2k_Fx)]e^{i\theta(x)}\\
&\hspace{0.5cm}+[C_3(l)+C_4(l)\cos(2\phi(x)-2k_Fx)]e^{-2i\theta(x)},
\end{split}
\end{equation}
with $C_1(0) = C_2(0) = 1/\sqrt{2\pi\alpha}$, and $C_3(0)=C_4(0)=0$. Terms with larger scaling dimensions can be neglected in describing the behavior of the correlation functions. Since the bosonic fields in Eq.\,\eqref{hard-core_ansatz} are those of the original Hamiltonian \eqref{3boson_bosonization}, 
the flow of the rescaled operator need to be considered. We calculate here the flow of $e^{i\theta(x)}$, from which the flow equation for $C_2(l)$ can be deduced. We have
\begin{equation}\label{b_vertex}
e^{i\theta(x)/\sqrt{K}}=\Big(\frac{2\pi\alpha}{L}\Big)^{\beta^2/36}V_R(-\beta/6,x)V_L(-\beta/6,x).
\end{equation} 
The details of the calculation can be found in Appendix \ref{appendix:Calculations_details}. We summarize here the main steps of the derivation. First, the commutator of Eq.\,\eqref{b_vertex} with the hermitian conjugate part of the generator \eqref{generator_dualsG} leads to less relevant terms that are truncated in the ansatz \eqref{hard-core_ansatz}. The most relevant operator in the remaining part of $[\eta_{\text{dual}}^{(1)}(l),e^{i\theta(x)/\sqrt{K}}]$ is extracted by an OPE of the vertex operators. The parameters of the vertex operators in Eqs.\,\eqref{generator_dualsG} and \eqref{b_vertex} combine to produce the operator $e^{-2i\theta(x)}$ in the ansatz \eqref{hard-core_ansatz}. Namely,
\begin{equation}
e^{-2i\theta(x)/\sqrt{K}}=\Big(\frac{2\pi\alpha}{L}\Big)^{\beta^2/9}V_R(\beta/3,x)V_L(\beta/3,x).
\end{equation}
The flow equation for $C_3(l)$ is then given by
\begin{equation}\label{flow_C2}
\frac{dC_3(l)}{dl} = -C_1(l)\frac{4v_Fg(l)}{\Gamma(\beta^2/12)^2}\frac{2\pi}{L}\sum_{k>0}k|\alpha k|^{\beta^2/6-2}e^{-4v_F^2k^2l}.
\end{equation}
To the leading order in the bare coupling constant $g$, we can replace $C_1(l)$ by its bare value. Moreover, the running coupling constant can be replaced by its approximate solution in the weak-coupling regime\,\cite{Sabio_Kehrein_2010}. From the RG equations \eqref{RG_eq}, we have
\begin{equation}
g(l_{\text{RG}}) \sim ge^{(2-\beta^2/4)l_{\text{RG}}} = g\Big(\frac{32l}{\alpha^2}\Big)^{1-\beta^2/8}.
\end{equation}
Finally, the solution of Eq.\,\eqref{flow_C2} at the end of the flow reads
\begin{equation}
C_3(\infty) = -\frac{4v_F g}{\sqrt{2\pi\alpha}}D_{\beta}\int_{0}^{\infty} dk\,k^{\beta^2/6-1}f(2v_F k),
\end{equation}
with $D_{\beta} = \frac{(32)^{1-\beta^2/8}}{\Gamma(\beta^2/12)}$ and $f(k)=k^{\beta^2/4-4}\Gamma(2-\beta^2/8,k^2)$.
\subsubsection{Bosonic correlation functions}
\begin{figure}[H]
 \centering
  \includegraphics[scale=0.5]{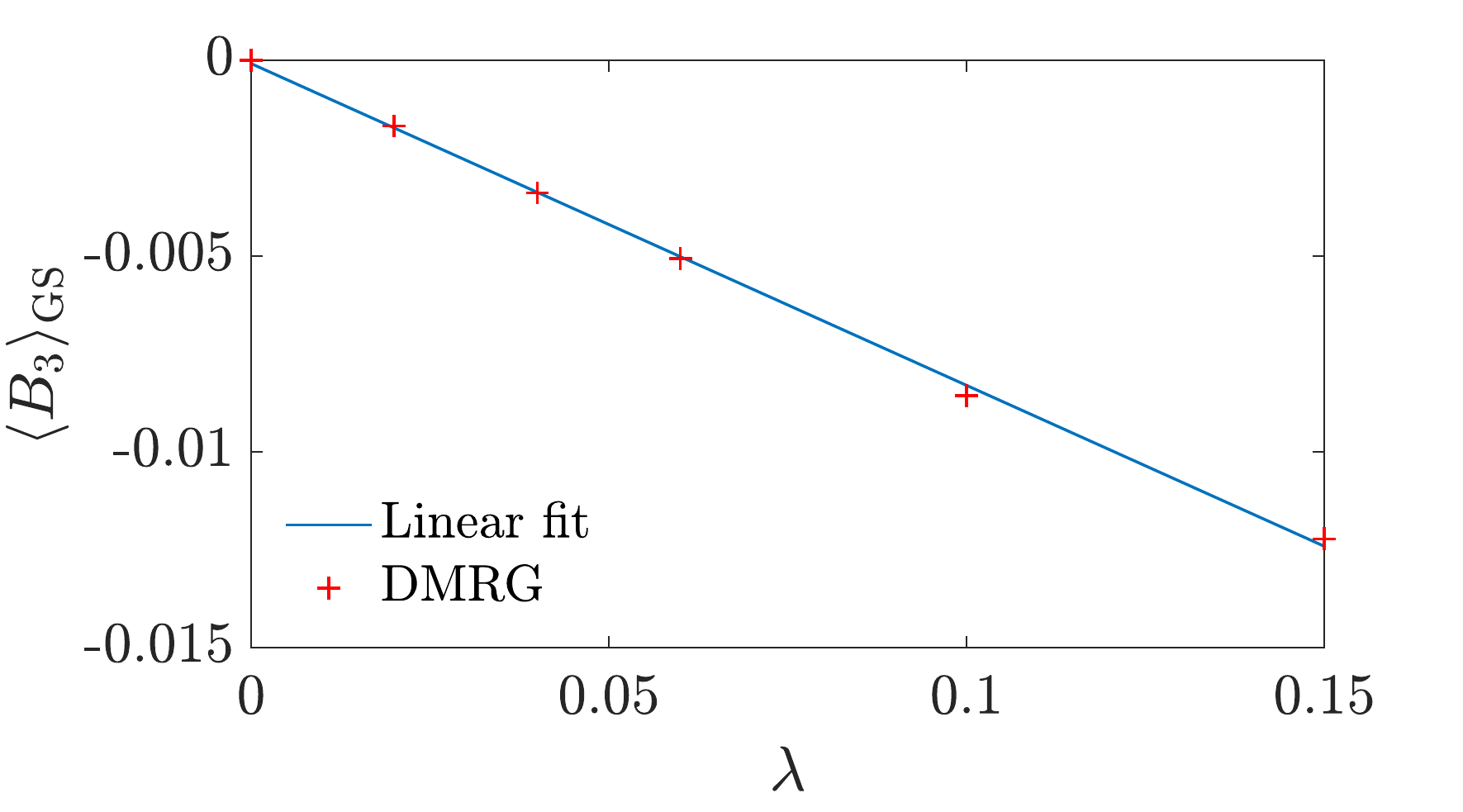}
 \caption{DMRG calculations of the expectation value of the three-boson operator $B_3$, in the ground state of the bare Hamiltonian \eqref{3boson_model}, as a function of the three-site interaction coupling $\lambda$ and at $\mu=1$.}
 \label{B3vslambda}
\end{figure}

 We consider the correlation functions of the point-split product of $p$ hard-core bosonic operators
\begin{equation}\label{product_bosons}
B_p(x) = \lim_{\Delta\to0}\prod_{n=0}^{p-1}b(x+n\Delta).
\end{equation}
As in section \ref{section:transformed_fermion}, the correlation functions in the ground state of the bare Hamiltonian \eqref{3boson_model} can be evaluated from the bosonic representation of the transformed operator $\Tilde{B}_p(x)=U(\infty)B_p(x)U^\dagger(\infty)$.  It is derived from $B_p(x)$ by substituting the hard-core boson operators with their transformed counterpart $b(x,\infty)$. The one-boson and two-boson operators are  given by
\begin{equation}\label{relation}
\begin{split}
\Tilde{B}_1(x) &\sim C_1e^{i\theta(x)}+C_3e^{-2i\theta(x)},\\
\Tilde{B}_2(x) &\sim (C_1)^2e^{2i\theta(x)}+2C_1C_3e^{-i\theta(x)},
 \end{split}
\end{equation}
where the coefficients are taken at the end of the flow, i.e. at $l=\infty$. The oscillating terms in Eq.\,\eqref{hard-core_ansatz} are neglected since they have larger scaling dimensions.
Using the result \eqref{correlation_LL} for the correlation functions in the Luttinger liquid Hamiltonian, we obtain
\begin{align}\label{relation}
 \expval{B^\dagger_1(x)B_1(y)}_{\text{GS}} &\sim (C_1)^2\frac{1}{r^{\frac{1}{2K}}}+(C_3)^2\frac{1}{r^{\frac{2}{K}}},\\
 \expval{B^\dagger_2(x)B_2(y)}_{\text{GS}} &\sim (C_1)^4\frac{1}{r^{\frac{2}{K}}}+4(C_1C_3)^2\frac{1}{r^{\frac{1}{2K}}},
\end{align}
with $r=|x-y|$. 
For $p=3$, a combination of $e^{i\theta(x)}$ in the product \eqref{product_bosons} can lead to a vanishing exponent. In this case, a higher-order expansion in the point-splitting parameter $\Delta$ need to be carried out. Moreover, the Klein factors become necessary as their anti-commutation prevents artificial cancellations. They are reintroduced in the ansatz by replacing $\cos(2\phi)$ with $F_Re^{2i\phi}+F_Le^{-2i\phi}$. The details of the derivation are presented in Appendix \ref{appendix:Calculations_details}. Finally, the correlation function takes the form
\begin{equation}\label{3bosons_corr}
\begin{split}
\expval{B^\dagger_3(x)B_3(y)}_{\text{GS}} \sim& D_0\frac{1}{r^{\frac{9}{2K}}}+D_1\frac{1}{r^2}+D_2\frac{\cos(2k_Fr)}{r^{2K}}\\
&+D_3\frac{\cos(2k_Fr)}{r^{2K+2}}+D_4\frac{1}{r^4},
\end{split}
\end{equation}
where $D_1$, $D_2$, $D_3$ and $D_4$ depend on the coefficient of the transformed operator $b(x,\infty)$.
\begin{figure}[H]
 \centering
  \includegraphics[width=0.95\textwidth]{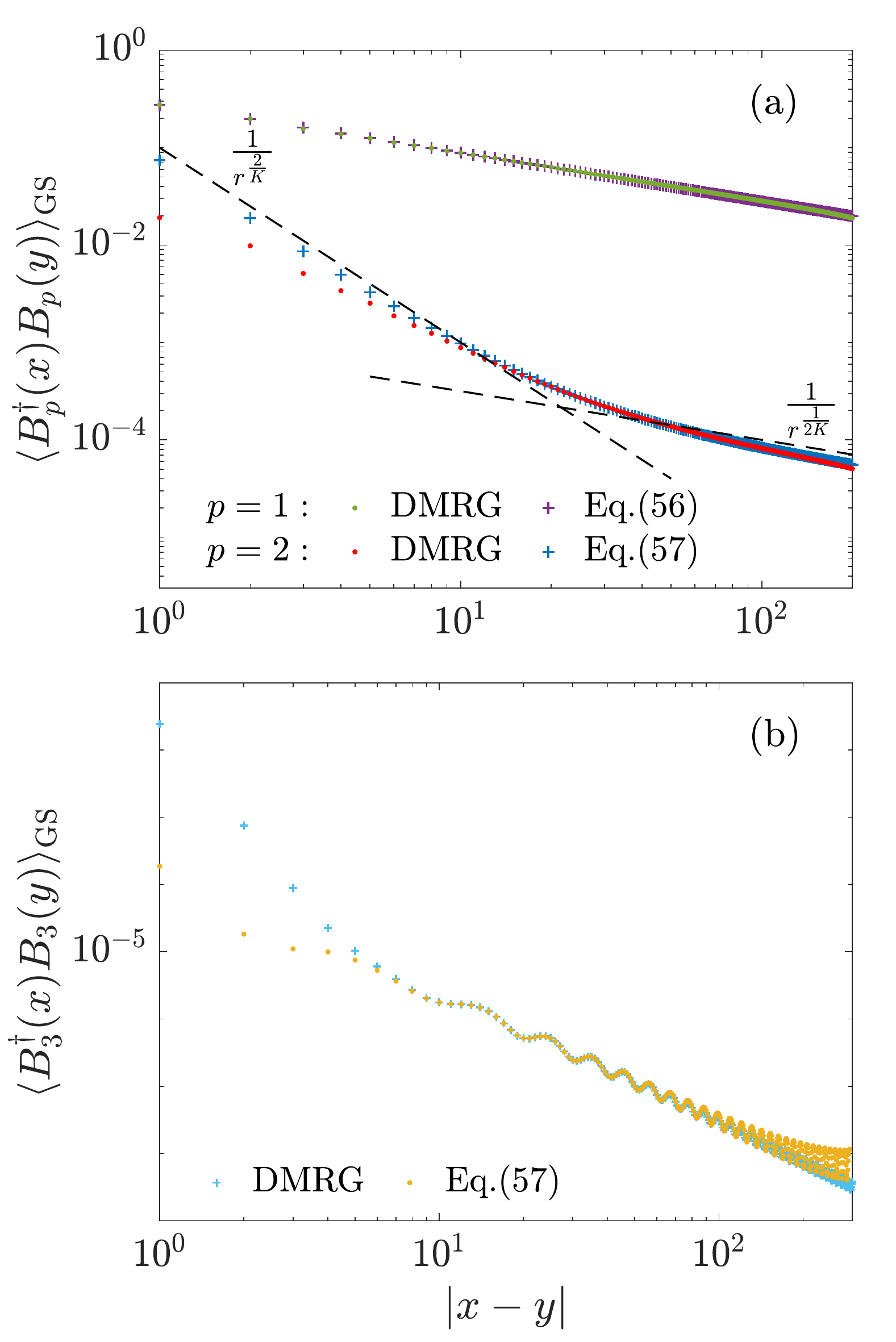}
 \caption{(a) One-boson and two-boson correlation functions at $\mu=1$, $\lambda=0.06$ and (b) Connected three-boson correlation functions at $\mu=1.9$, $\lambda=0.2$, in the ground state of the bare Hamiltonian. The dashed line are provided as a guide to the eye. The DMRG calculations are performed on a system of size $N=1200$. The Luttinger parameter $K$ and the Fermi wave vector $k_F$ are extracted from Friedel oscillations as discussed in section \ref{section:bosonization} and the coefficients in the correlations are obtained from a least square fit.}
 \label{DMRG_correlation_bosons}
\end{figure} 
As shown in Fig.\ref{DMRG_correlation_bosons}, the correlations decay algebraically at short distance with the exponent $p^2/2K$, associated with the operator $e^{ip\theta(x)}$ in the standard bosonic representation of the product of $p$ hard-core bosons. At large distances, the two-boson and three-boson correlations undergo a crossover to power-laws with smaller exponents, induced by the terms generated along the flow. In particular, the two-boson correlations exhibit a crossover to the one-boson correlations. Similarly, the three-boson correlations acquire, among other terms, an oscillating part with a wave vector $2k_F$ that decays with an exponent $2K$. We also note that the presence of gradients of bosonic fields in $\Tilde{B}_3(x)$ leads to a non-vanishing expectation of the three-boson operator inside the Luttinger liquid phase. Since these terms are generated to the first order of the coupling $g\propto\lambda$, the expectation decays linearly with the interaction strength and vanish at the non-interacting point $\lambda=0$. This result is confirmed by the numerical simulations (Fig.\ref{B3vslambda}).

\section{Conclusion}\label{section:Conclusion}
Using the flow equation approach, we have provided an analytical derivation of modified bosonic representations of hard-core boson and spinless fermion operators that take account of the $\mathbb{Z}_3$ symmetry of the Hamiltonian. As a result, the correlation functions of $p$-particles are dominated in the long distances by the power-law decay of operators that are not initially present in the bosonic representation of the operators. These calculations can be straightforwardly generalized to $\mathbb{Z}_n$-symmetric models. Since the generator of the flow exhibits the symmetry of the bare Hamiltonian by construction, terms that transform covariantly under $\mathbb{Z}_n$-rotations will be generated in the bosonic representation of single-particle operators. For instance, in the hard-core boson model with creation/annihilation operators on four adjacent sites, which is of interest in the open problem of commensurate melting of density-waves, the bosonic representation of a product of $p$ bosons is expected to contain terms that are associated with $p-4m$ particles, where $m\in \mathbb{Z}$. Some of these terms have smaller scaling dimension than the bare operator and will dominate the correlations at long distances.

More generally, this paper demonstrates that even if an emergent U(1) symmetry is present, the long distance correlation functions can have a behaviour that differs from the naive expectation and that the flow equation approach is a very useful tool to identify the correct bosonic representation of lattice operators. It would be interesting to investigate the extension of this method to models with an emergent non-abelian symmetry.
\section*{Acknowledgments}
The work has been supported by the Swiss National Science Foundation (FM, ZJ) grant 212082 and by the Delft Technology Fellowship (NC). Numerical simulations have been performed on the Dutch national e-infrastructure with the support of the SURF Cooperative and the support of the Scientific IT and the Application Support Center of EPFL. We thank Melvyn Nabavi for proof reading the manuscript.
\newpage
\appendix
\onecolumngrid
\renewcommand{\appendixname}{APPENDIX}

\section{\uppercase{Derivation of the flow equations}}
\label{appendix:Derivation_floweq}
\subsection{Flow equations for the fermionic Hamiltonian}
An efficient way of computing the flow equations is to use Wick's theorem for operator products to collect the contributions to the different generated terms in  $[\eta,H_3]$. For convenience, we write the interaction term as 
\begin{equation}
H_{3} = \frac{1}{3!\sqrt{N}}\sum_{\vec{k}}\Tilde{B}_{k_1,k_2,k_3}\Big(c^\dagger_{k_1}c^\dagger_{k_2}c^\dagger_{k_3}+c_{k_1}c_{k_2}c_{k_3}\Big),
\end{equation}
where $\vec{k} = (k_1, k_2, k_3)$ and
\begin{equation}
\Tilde{B}_{k_1,k_2,k_3} = \delta_{k_1+k_2+k_3,0}\sum_{\sigma \in S_{3}}\epsilon(\sigma)e^{-ik_{\sigma(2)}}e^{-2ik_{\sigma(3)}},
\end{equation}
is purely imaginary. $S_3$ is the symmetric group of order 3, and $\epsilon(\sigma)$ is the sign of the permutation. The contribution from $[\eta,H_3]$ to the flow of $U_{k,q,p}$ and $\xi_{k}$ stems from the following term:
\begin{equation}\label{etaH_3_Uxi}
\begin{split}
[\eta,H_3] \to& \frac{1}{(3!)^2 N}\sum_{\vec{k}}\Tilde{B}_{k_1,k_2,k_3}\Tilde{B}_{q_1,q_2,q_3}\Tilde{\alpha}_{k_1,k_2,k_3}\Big([c^\dagger_{k_1}c^\dagger_{k_2}c^\dagger_{k_3},c_{q_1}c_{q_2}c_{q_3}]+(\vec{k}\leftrightarrow\vec{q})\Big),
\end{split}
\end{equation}
where $\Tilde{\alpha}_{k_1,k_2,k_3}=\sum_{i=1}^{3}\xi_{k_i}$. To proceed with the calculation, it is useful to apply Wick's theorem in order to normal order the generated terms with respect to the Fermi sea in the ground state of the free Hamiltonian $H_0$. The contractions of two-fermion operators are given by

\begin{equation}
\wick{\c1c^\dagger_{k} \c1c_{q}} = \Theta(k_F-|k|)\delta_{k,q},\,\,\wick{\c1c_{k} \c1c^\dagger_{q}} = [1-\Theta(k_F-|k|)]\delta_{k,q}.
\end{equation}

To compute the flow equation of the two-body interactions, we need to collect all the normal-ordered terms that result from single contractions in Eq.\eqref{etaH_3_Uxi}. These are given by
\begin{equation}\label{single_contraction}
\begin{split}
:c^\dagger_{k_1}c^\dagger_{k_2}c^\dagger_{k_3}::c_{q_1}c_{q_2}c_{q_3}:\,\,&\to \Theta(k_F-|k_1|)\delta_{k_1,q_1}:c^\dagger_{k_2}c^\dagger_{k_3}c_{q_2}c_{q_3}:+\text{other single contractions}\\
&= \Big(\frac{1}{2!}\Big)^2\sum_{\sigma,\sigma'\in S_3}\epsilon(\sigma)\epsilon(\sigma')\Theta(k_F-|k_{\sigma(1)}|)\delta_{k_{\sigma(1)},q_{\sigma'(1)}}:c^\dagger_{k_{\sigma(2)}}c^\dagger_{k_{\sigma(3)}}c_{q_{\sigma'(2)}}c_{q_{\sigma'(3)}}:,\\
\end{split}
\end{equation}
where the columns denote the normal-ordering with respect to the Fermi sea and the factor $(1/2!)^2$ avoids over-counting contractions. Similarly, 
\begin{equation}
:c_{q_1}c_{q_2}c_{q_3}::c^\dagger_{k_1}c^\dagger_{k_2}c^\dagger_{k_3}:\,\,\to\Big(\frac{1}{2!}\Big)^2\sum_{\sigma,\sigma'\in S_3}\epsilon(\sigma)\epsilon(\sigma')[1-\Theta(k_F-|k_{\sigma(1)}|)]\delta_{k_{\sigma(1)},q_{\sigma'(1)}}:c^\dagger_{k_{\sigma(2)}}c^\dagger_{k_{\sigma(3)}}c_{q_{\sigma'(2)}}c_{q_{\sigma'(3)}}:.
\end{equation}
Therefore, the contribution to the two-body interaction term reads 
\begin{equation}\label{etaH_3_U2}
\begin{split}
[\eta,\lambda H_3] \to& \Big(\frac{1}{2!}\Big)^2\frac{\lambda^2}{(3!)^2 N}\sum_{\vec{k},\vec{q}}\Tilde{B}_{k_1,k_2,k_3}\Tilde{B}_{q1,q_2,q_3}(\Tilde{\alpha}_{k_1,k_2,k_3}+\Tilde{\alpha}_{q_1,q_2,q_3})\\
&\hspace{5cm}\times\sum_{\sigma,\sigma'\in S_3}\epsilon(\sigma)\epsilon(\sigma')[2\Theta(k_F-|k_{\sigma(1)}|)-1]\delta_{k_{\sigma(1)},q_{\sigma'(1)}}:c^\dagger_{k_{\sigma(2)}}c^\dagger_{k_{\sigma(3)}}c_{q_{\sigma'(2)}}c_{q_{\sigma'(3)}}:\\
=& \Big(\frac{1}{2!}\Big)^2\frac{1}{(3!)^2 N}\sum_{\vec{k},\vec{q}}\Tilde{B}_{k_{\sigma^{-1}(1)},k_{\sigma^{-1}(2)},k_{\sigma^{-1}(3)}}\Tilde{B}_{q_{\sigma'^{-1}(1)},q_{\sigma'^{-1}(2)},q_{\sigma'^{-1}(3)}}(\Tilde{\alpha}_{k_{\sigma^{-1}(1)},k_{\sigma^{-1}(2)},k_{\sigma^{-1}(3)}}+\Tilde{\alpha}_{q_{\sigma'^{-1}(1)},q_{\sigma'^{-1}(2)},q_{\sigma'^{-1}(3)}})\\
&\hspace{5cm}\times\sum_{\sigma,\sigma'\in S_3}\epsilon(\sigma)\epsilon(\sigma')[2\Theta(k_F-|k_{1}|)-1]\delta_{k_{1},q_{1}}:c^\dagger_{k_{2}}c^\dagger_{k_{3}}c_{q_{2}}c_{q_{3}}:\\
=& \Big(\frac{3!}{2!}\Big)^2\frac{1}{(3!)^2 N}\sum_{\vec{k},\vec{q}}\Tilde{B}_{k_1,k_2,k_3}\Tilde{B}_{	k_1,q_2,q_3}(\Tilde{\alpha}_{k_1,k_2,k_3}+\Tilde{\alpha}_{k_1,q_2,q_3})[2\Theta(k_F-|k_{1}|)-1]:c^\dagger_{k_{2}}c^\dagger_{k_{3}}c_{q_{2}}c_{q_{3}}:.\\
\end{split}
\end{equation}
In the second line, the change of indices $k_{\sigma'(i)}\to k_i$ for $i=1,2,3$ is made in order to shift the permutations dependence to the indices of the interaction matrix. Finally, the last line is obtained from the anti-symmetry of the interaction matrix under permutation, i.e., $B_{k_{\sigma(1)},k_{\sigma(2)},k_{\sigma(3)}}=\epsilon(\sigma)B_{k_1,k_2,k_3}$. A similar result can be obtained for terms generated from double contractions, which contribute to the dispersion $\xi_k$. After writing Eq.\,\eqref{etaH_3_U2} in terms of $B_{k,q}=\Tilde{B}_{k,q,-k-q}$ and $\alpha_{k,q}=\Tilde{\alpha}_{k,q,-k-q}$, the contribution to the two-body term reads
\begin{equation}
[\eta,\lambda H_3]\to\frac{\lambda^2}{4N}\sum_{k,q,p} B_{k+p,q-p}B_{q,k}(\alpha_{k+p,q-p}+\alpha_{k,q})[2\Theta(k_F-|k+q|)-1]:c^\dagger_{k+p}c^\dagger_{q-p}c_{q}c_{k}:.\\
\end{equation}
Thus, the flow equation of the two-body interaction is given by 
\begin{equation}
\frac{dU_{k,q,p}(l)}{dl} = -\frac{1}{4} B_{k+p,q-p}(l)B_{k,q}(l)(\alpha_{k+p,q-p}(l)+\alpha_{k,q}(l))[2\Theta(k_F-|k+q|)-1].
\end{equation} 
\subsection{Flow equations for the hard-core bosonic Hamiltonian}
The flow equations of the sine-Gordon model are derived in Ref.\,\onlinecite{Kehrein_2001}. We have
\begin{equation}
\begin{split}
\frac{d\Tilde{\beta^2}(l_{\text{RG}})}{dl_{\text{RG}}}&=-\frac{\Tilde{\beta}^4(l_{\text{RG}})}{4\Gamma(\frac{\Tilde{\beta}^2(l_{\text{RG}})}{4}-1)}g^2(l_{\text{RG}}),\\
\frac{dg(l_{\text{RG}})}{dl_{\text{RG}}}&=\Big(2-\frac{\Tilde{\beta}^2(l_{\text{RG}})}{4}\Big)g(l_{\text{RG}}).
\end{split}
\end{equation}
Eqs.\,\eqref{floweq_dualsG} are deduced from the relation $\Tilde{\beta}=3\sqrt{K}$ and the duality transformation $K\to 1/K$. 
\section{\uppercase{Bosonization dictionary}}\label{appendix:Bosonization_dic}
\subsection{Bosonic fields}
The bosonic field $\phi$, $\theta$ are constructed from the density modes $\rho^\dagger_{r}(p)$, where $r=R,L$ denotes their species. We have
\begin{equation}\label{boz_fields}
\begin{split}
\phi(x) = -\frac{i\pi}{L}\sum_{p\neq0}\frac{e^{-\alpha|p|/2-ipx}}{p}\big[\rho_R^\dagger(p)+\rho_L^\dagger(p)\big],\\
\theta(x) =  \frac{i\pi}{L}\sum_{p\neq0}\frac{e^{-\alpha|p|/2-ipx}}{p}\big[\rho^\dagger_R(p)-\rho^\dagger_L(p)\big],
\end{split}
\end{equation}
where
\begin{equation}
[\rho_r^\dagger(p), \rho_{r'}^\dagger(-q)] = -\delta_{r,r'}\delta_{p,q}\frac{rpL}{2\pi}.
\end{equation}
The relation between the density modes and the normal modes of the sine-Gordon model is given by
\begin{equation}
\begin{split}
\rho^\dagger_R(p) =\sqrt{|p|}\sigma_1(p),\\
\rho^\dagger_L(p)=\sqrt{|p|}\sigma_2(p).
\end{split}
\end{equation}
The relation between the fields in Eq.\,\eqref{boz_fields} and the fields $\Tilde{\phi}$ and $\Tilde{\theta}$ introduced in the reference is given by
\begin{equation}
\begin{split}
\phi(x) = \sqrt{\pi}\Tilde{\phi}(x),\\
\theta(x) = -\sqrt{\pi}\Tilde{\theta}(x).
\end{split}
\end{equation}
\subsection{Vertex operators}
We define the vertex operators as 
\begin{equation}\label{vertex}
V_{r}(\beta,x) =\,:e^{i\beta[r\phi(x)-\theta(x)]}:.
\end{equation}
Note that this definition coincides with the vertex operators $\Tilde{V}_r(\beta,x) =\,:e^{i\sqrt{\pi}\beta[r\Tilde{\phi}(x)-\Tilde{\theta}(x)]}:$ introduced in the reference. In terms of the density modes, Eq.\,\eqref{vertex} reads
\begin{equation}
V_r(\beta,x) =\,:\exp(r\beta\frac{2\pi}{L}\sum_{p\neq0}\frac{e^{-\alpha|p|/2-ipx}}{p}\rho^\dagger_{r}(p)):.
\end{equation} 
The normal ordering with respect to the vacuum defined by
\begin{equation}
\begin{split}
\rho^\dagger_R(p<0)\ket{0} = 0,\\
\rho^\dagger_L(p>0)\ket{0} = 0,
\end{split}
\end{equation}
yields
\begin{equation}
V_{r}(\beta,x) = \Big(\frac{L}{2\pi\alpha}\Big)^{\beta^2/2}e^{i\beta[r\phi(x)-\theta(x)]}.
\end{equation}
The bosonization mapping is then given by
\begin{equation}
\psi_r(x) = \frac{e^{irk_Fx}}{\sqrt{L}}V_r(-1,x).
\end{equation}
\subsubsection{Operator product expansion}
The operator product expansion of a product of vertex operators is given by
\begin{equation}
\begin{split}
V_R(\beta,x)V_R(-\gamma,y) \sim \Big(\frac{L/2\pi}{i(y-x)+\alpha}\Big)^{\beta\gamma} V_R(\beta-\gamma,x),\\
V_L(\beta,x)V_L(-\gamma,y) \sim \Big(\frac{L/2\pi}{i(x-y)+\alpha}\Big)^{\beta\gamma} V_L(\beta-\gamma,x).
\end{split}
\end{equation}
\subsubsection{Exchange relations}
The order of vertex operators can be exchanged using the following relations:
\begin{equation}
    \begin{split}
    V_{R}(-\gamma,y)V_{R}(\beta,x) &\sim V_{R}(\beta,x)V_{R}(-\gamma,y)\times\frac{[i(y-x)+\alpha]^{\beta\gamma}}{[i(x-y)+\alpha]^{\beta\gamma}},\\
     V_{L}(-\gamma,y)V_{L}(\beta,x) &\sim V_{L}(\beta,x)V_{L}(-\gamma,y)\times\frac{[i(x-y)+\alpha]^{\beta\gamma}}{[i(y-x)+\alpha]^{\beta\gamma}},
    \end{split}
\end{equation}
and $[V_R(\gamma,x),V_L(\delta,y)]=0$ for all $\gamma, \delta$. 
\subsubsection{Solition and antisoliton operators}
The soliton and antisoliton operators in the effective Hamiltonian \eqref{H_d} are defined as the Fourier transform of the vertex operators:
\begin{equation}
\Tilde{P}_r(k) = \Bigg[\frac{\Gamma(\Tilde{\beta}^2/4)}{2\pi L}\Big(\frac{L|k|}{2\pi}\Big)^{1-\frac{\Tilde{\beta}^2}{4}}\Bigg]^{1/2}\int dx\,e^{-ikx}V_r(-\Tilde{\beta}/2,x),\,\,\,\text{for $r=R,L$}.
\end{equation}

\section{\uppercase{Derivation of the Luttinger liquid Hamiltonian}}\label{appendix:Derivation_K}
We derive here the Luttinger parameter and the velocity inside the Luttinger liquid phase of the model \eqref{3fermion_model}. We consider the particle-hole excitations close to the Fermi point $\pm k_F$ in the two-body interactions of the effective Hamiltonian $H_{\text{eff}}=H_0(\infty)+\lambda^2 H_U(\infty)$. They consist in two $g_2$ processes and two $g_1$ processes. Since the fermions are spinless, these two processes are indistinguishable and the two-body interaction term reduces to
\begin{equation}
H_{U}(\infty) = \frac{4g_2}{N}\sum_{p}\rho_R(p)\rho_L(-p),
\end{equation}
where 
\begin{equation}
g_2 = U_{k_F,-k_F,0} = -\frac{16\sin^2(k_F)\sin^4(k_F/2)}{2+4\cos(k_F)+3\mu},
\end{equation}
and 
\begin{equation}
\rho_{r}(p) = \sum_{k}c^\dagger_{r,k+p}c_{r,k}
\end{equation}
are the Fourier components of the density operators at the right ($r=R$) and left branches ($r=L$). Using the expressions of the density operators in terms of the fields $\phi$ and $\theta$:
\begin{equation}
\begin{split}
\rho_R(x) &= -\frac{1}{2\pi}\big[\partial_x\phi(x)-\partial_x\theta(x)\big],\\
\rho_L(x) &= -\frac{1}{2\pi}\big[\partial_x\phi(x)+\partial_x\theta(x)\big],
\end{split}
\end{equation}
the two-body interaction becomes
\begin{equation}\label{H_U_boz}
H_{U}(\infty) = \frac{4g_2}{(2\pi)^2}\int dx\,[\partial_x\phi(x)]^2-[\partial_x\theta(x)]^2.
\end{equation}
Eq.\,\eqref{H_U_boz} is then combined with the non-interacting Luttinger liquid Hamiltonian such that the Hamiltonian remains quadratic. We obtain
\begin{equation}
H_{U}(\infty) = \frac{u}{2\pi}\int dx\,K[\partial_x\phi(x)]^2+\frac{1}{K}[\partial_x\theta(x)]^2,
\end{equation} 
where
\begin{equation}\label{K_u}
\begin{split}
K &=\Big[\frac{1-2g_2\lambda^2/\pi\Tilde{v}_F}{1+2g_2\lambda^2/\pi\Tilde{v}_F}\Big]^{1/2},\\
u &= \Tilde{v}_F\Big[1-\Big(\frac{2g_2\lambda^2}{\pi\Tilde{v}_F}\Big)^2\Big]^{1/2},
\end{split}
\end{equation}
and $\Tilde{v}_F=\partial\xi_k/\partial k|_{k=k_F}$ is the renormalized velocity. Eq.\,\eqref{K_prediction} is obtained from an expansion of Eq.\,\eqref{K_u} to the second order of $\lambda$. 
\section{\uppercase{Some details of the calculations}}\label{appendix:Calculations_details}
\subsection{Derivation of the flow of the hard-core boson operator}\label{boson_flow_calculation}
We give here the detailed calculation of the flow equation for $C_3$ in Eq.\,\eqref{hard-core_ansatz}. We have
\begin{equation}\label{flow_expr}
[\eta_{\text{dual}}^{(1)}(l), e^{i\theta(x)/\sqrt{K}}] = -2iv_F\Big(\frac{2\pi\alpha}{L}\Big)^{\beta^2/36}\int dydz\frac{\partial u(z,l)}{\partial y}\big[V_R(\beta/2,y)V_L(\beta/2,y-z)+\text{h.c.}, V_R(-\beta/6,x)V_L(-\beta/6,x)\big].
\end{equation}
Since the terms obtained from the hermitian conjugation in Eq.\,\eqref{flow_expr} are less relevant terms, we only consider the commutator
\begin{equation}
\begin{split}
\big[V_R(\beta/2,y)V_L(\beta/2,y-z), V_R(-\beta/6,x)V_L(-\beta/6,x)\big] &=V_R(\beta/2,y) V_R(-\beta/6,x)V_L(\beta/2,y-z)V_L(-\beta/6,x)\\
&\times\Big\{1-\Big[\frac{i(x-y)+\alpha}{i(y-x)+\alpha}\Big]^{\beta^2/12}\Big[\frac{i(y-z-x)+\alpha}{i(x-y+z)+\alpha}\Big]^{\beta^2/12}\Big\}\\
&\hspace{-8cm}\sim \Big(\frac{L}{2\pi}\Big)^{\beta^2/6}V_R(\beta/3,x)V_L(\beta/3,x)\\
&\hspace{-6cm}\times\Big\{[i(x-y)+\alpha]^{-\beta^2/12}[i(y-z-x)+\alpha]^{-\beta^2/12}-[i(y-x)+\alpha]^{-\beta^2/12}[i(x-y+z)+\alpha]^{-\beta^2/12}\Big\},
\end{split}
\end{equation}
where an OPE of the vertex operator is carried out in last line. By inserting this expression in Eq.\,\eqref{flow_expr}, we obtain, after a few steps of calculations,
\begin{equation}
\begin{split}
&[\eta^{(1)}_{\text{dual}}(l),e^{i\theta(x)/\sqrt{K}}]\to -2iv_F\alpha^{\beta^2/6}\frac{g(l)}{4\pi^2\alpha^2}\frac{2\pi}{L}\sum_{k}(-ik)e^{-4v_F^2k^2l}e^{-2i\theta(x)/\sqrt{K}}\\
&\hspace{1cm}\times\int dydz\,e^{-ikz}\Big\{[i(x-y)+\alpha]^{-\beta^2/12}[i(y-z-x)+\alpha]^{-\beta^2/12}-[i(y-x)+\alpha]^{-\beta^2/12}[i(x-y+z)+\alpha]^{-\beta^2/12}\Big\}\\
&\hspace{2.9cm}= -\frac{v_Fg(l)}{\pi^2}\frac{2\pi}{L}\sum_k k e^{-4v_F^2k^2l}f_{\beta}(k)e^{-2i\theta(x)/\sqrt{K}},
\end{split}
\end{equation}
with 
\begin{equation}
\begin{split}
f_{\beta}(k) &= \int dydz\,e^{-ik\alpha z}[1-iy]^{-\beta^2/12}[1+i(y-z)]^{-\beta^2/12}\\
&= |\alpha k|^{\beta^2/6-2}\frac{4\pi^2}{\Gamma(\beta^2/12)^2}\Theta(k).
\end{split}
\end{equation}
Therefore, the flow equation for $C_3(l)$ is given by
\begin{equation}
\frac{dC_3(l)}{dl} = -C_1(l)\frac{4v_Fg(l)}{\Gamma(\beta^2/12)^2}\frac{2\pi}{L}\sum_{k>0}k|\alpha k|^{\beta^2/6-2}e^{-4v_F^2k^2l}.
\end{equation}

\subsection{Derivation of $\Tilde{B}_3(x)$}\label{appendix:operators_calculation}
We give here the detailed calculation of the three-boson correlations in Eq.\,\eqref{3bosons_corr}. We start by reordering the vertex operators in the product $\Tilde{B}_3(x)$. This leads to
\begin{equation}\label{TildeB_3}
\begin{split}
\Tilde{B_3}(x)&= \lim_{\Delta\to 0}\sum_{n,m,l=0,1}\Big[A_n+B_n\cos(2\Tilde{\phi}(x))\Big]\Big[A_m+B_m(-1)^{p_n}\cos(2\Tilde{\phi}(x+\Delta))\Big]\\
&\hspace{4cm}\times\Big[A_l+B_l(-1)^{p_n+p_m}\cos(2\Tilde{\phi}(x+2\Delta))\Big]e^{i[p_n\theta(x)+p_m\theta(x+\Delta)+p_l\theta(x+2\Delta)]},\\
\end{split}
\end{equation}
where $\Tilde{\phi}(x)=\phi(x)-k_Fx$, $p_n=1-3n$, $A_0=C_0$, $B_0=C_1$, $A_1=C_3$ and $B_1=C_4$. To avoid artificial cancellations, it is necessary to reintroduce the Klein factors in the definition of the ansatz. This amounts to the following substitution:
\begin{equation}
\begin{split}
B_n\cos(2\phi(x))&\to \frac{1}{2}B_n\big[F_Re^{2i\phi(x)}+F_Le^{-2i\phi(x)}\big],\\
A_n &\to \frac{1}{2}A_n(F_R+F_L).
\end{split}
\end{equation}
By carrying out an expansion in the splitting parameter $\Delta$, we obtain for a term at position $x+(n-1)\Delta$ in the product of Eq.\,\eqref{TildeB_3}:
\begin{equation}
\begin{split}
&\frac{1}{2}\big[F_Re^{2i\phi(x+(n-1)\Delta)}+F_Le^{-2i\phi(x+(n-1)\Delta)}\big]\\
&\hspace{2cm}\sim \frac{1}{2}F_R\Big\{1+2i[(n-1)\Delta\partial_x\phi+\frac{(n-1)^2\Delta^2}{2}(\partial_x\phi)^2]-2(n-1)^2\Delta^2\partial_x^2\phi\Big\}e^{2i\phi}\\
&\hspace{2.5cm}+\frac{1}{2}F_L\Big\{1-2i[(n-1)\Delta\partial_x\phi+\frac{(n-1)^2\Delta^2}{2}(\partial_x\phi)^2]-2(n-1)^2\Delta^2\partial_x^2\phi\Big\}e^{-2i\phi},
\end{split}
\end{equation}
where the position of the fields is at $x$ and is omitted for brevity.  
Similarly, we have
\begin{equation}
\begin{split}
e^{i[p_n\theta(x)+p_m\theta(x+\Delta)+p_l\theta(x+2\Delta)]} \sim e^{i[p_n+p_m+p_l]\theta}e^{i[(p_m+2p_l)\Delta\partial_x\theta+(\frac{1}{2}p_m+2p_l)\Delta^2\partial_x^2\theta]}\\
\sim  e^{i[p_n+p_m+p_l]\theta}\big[1+i[(p_m+2p_l)\Delta\partial_x\theta+(\frac{1}{2}p_m+2p_l)\Delta^2\partial_x^2\theta]-\frac{1}{2}(p_m+2p_l)^2\Delta^2(\partial_x\theta)^2\Big].
\end{split}
\end{equation}
After collecting all the terms that have a scaling dimension smaller than the bare operator $e^{3i\theta(x)}$, we obtain
\begin{equation}\label{TildeB_3_2}
\begin{split}
\Tilde{B}_3(x) \sim &(C_0)^3 e^{3i\theta(x)}+[8(C_1)^2 C_4-(C_2)^2 C_4]\cos(2\Tilde{\phi})-12\Delta C_1C_2C_4\partial_x\theta\\
&+ \Delta [32C_1C_2C_3-8(C_1)^2C_4+(C_2)^2C_4]\partial_x\phi\sin(2\Tilde{\phi})-72\Delta^2(C_1)^2C_3(\partial_x\theta)^2\\
&+i\Delta^2[156(C_1)^2C_3-2(C_2)^2C_3+12C_1C_2C_4]\partial^2_x\theta+16\Delta^2(C_2)^2C_3(\partial_x\phi)^2.
\end{split}
\end{equation}
It should be noted that for simplicity of the calculation, the normal-ordering of the operators in Eq.\,\eqref{TildeB_3} is not taken before carrying out the Taylor expansion. The latter modifies the prefactors of the generated terms in Eq.\,\eqref{TildeB_3_2}. 
\twocolumngrid
\bibliographystyle{apsrev4-1}
\bibliography{bibliography, comments}
\end{document}